\documentclass{ar2e}
\usepackage{ulem}  
\usepackage{epsfig}

\begin{document}

\jname{Annu. Rev. Astron. Astrophys.}
\jyear{2010}
\jvol{48}
\ARinfo{1056-8700/97/0610-00}

\title{THE HUBBLE CONSTANT}

\markboth{Wendy L. Freedman \& Barry F. Madore}{Annu. Rev. Astron.
Astrophys. Vol. 48, 2010}

\author{Wendy L. Freedman and Barry F. Madore
\affiliation{Carnegie Observatories, 813 Santa Barbara St.,
Pasadena,  CA 91101, USA; email: wendy@obs.carnegiescience.edu, barry@obs.carnegiescience.edu}}


\begin{abstract}
Considerable progress has been made in determining the Hubble constant
over the past two decades.  We discuss the cosmological context and
importance of an accurate measurement of the Hubble constant, and
focus on six high-precision distance-determination methods: Cepheids,
tip of the red giant branch, maser galaxies, surface-brightnes
fluctuations, the Tully-Fisher relation and Type Ia supernovae. We
discuss in detail known systematic errors in the measurement of galaxy
distances and how to minimize them. Our best current estimate of the
Hubble constant is 73 $\pm$2 (random) $\pm$4
(systematic)~km~s$^{-1}$~Mpc$^{-1}$. The importance of improved
accuracy in the Hubble constant will increase over the next decade
with new missions and experiments designed to increase the precision
in other cosmological parameters. We outline the steps that will be
required to deliver a value of the Hubble constant to 2\% systematic
uncertainty and discuss the constraints on other cosmological
parameters that will then be possible with such accuracy.
\end{abstract}

\maketitle

\section{INTRODUCTION}

In 1929 Carnegie astronomer, Edwin Hubble, published a linear
correlation between the apparent distances to galaxies and their
recessional velocities. This simple plot provided evidence that our
Universe is in a state of expansion, a discovery that still stands as
one the most profound of the twentieth century (Hubble 1929a).  This
result had been anticipated earlier by Lema{\^i}tre (1927), who first
provided a mathematical solution for an expanding universe, and noted
that it provided a natural explanation for the observed receding
velocities of galaxies.  These results were published in the Annals
of the Scientific Society of Brussels (in French), and were not widely
known.

Using photographic data obtained at the 100-inch Hooker telescope
situated at Mount Wilson CA, Hubble measured the distances to six
galaxies in the Local Group using the Period-Luminosity relation
(hereafter, the Leavitt Law) for Cepheid variables. He then extended
the sample to an additional 18 galaxies reaching as far as the Virgo
cluster, assuming a constant upper limit to the brightest blue stars
(HII regions) in these galaxies. Combining these distances with
published radial velocity measurements (corrected for solar motion)
Hubble constructed Figure 1. The slope of the velocity versus distance
relation yields the Hubble constant, which parameterizes the current
expansion rate of the Universe.

The Hubble constant is usually expressed in units of kilometers per
second per megaparsec, and sets the cosmic distance scale for the
present Universe.  The inverse of the Hubble constant has dimensions
of time.  Locally, the Hubble law relates the distance to an object
and its redshift: cz = H$_0$d, where d is the distance to the object
and z is its redshift.  The Hubble law relating the distance and the
redshift holds in any Friedman-Lemaitre-Robertson-Walker cosmology
(see \S \ref{sec:cosmology}) for redshifts less than unity.  At
greater redshifts, the distance-redshift relationship for such a
cosmology also depends on the energy densities of matter and dark
energy.  The exact relation between the expansion age and the Hubble
constant depends on the nature of the mass-energy content of the
Universe, as discussed further in \S \ref{sec:cosmology} and \S
\ref{sec:age}. In a uniformly expanding universe, the Hubble
parameter, H(t), changes as a function of time; H$_\circ$, referred to
as the Hubble constant, is the value at the current time, $t_\circ$.

Measurement of the Hubble constant has been an active subject since
Hubble's original measurements of the distances to galaxies: the
deceptively simple correlation between galaxy distance and recession
velocity discovered eighty years ago did not foreshadow how much of a
challenge large systematic uncertainties would pose in obtaining an
accurate value for the Hubble constant.  Only recently have
improvements in linear, solid-state detectors, the launch of the
Hubble Space Telescope (HST), and the development of several different
methods for measuring distances led to a convergence on its current
value.  

Determining an accurate value for H$_o$ was one of the primary
motivations for building HST. In the early 1980's, the first director
of the Space Telescope Science Institute, Riccardo Giacconi, convened
a series of panels to propose observational programs of significant
impact requiring large amounts of Hubble observations. He was
concerned that in the course of any regular time allocation process
there would be reluctance to set aside sufficient time to complete
such large projects in a timely manner.  For decades a `factor-of-two'
controversy persisted, with values of the Hubble constant falling
between 50 and 100 km s$^{-1}$ Mpc$^{-1}$.  A goal of 10\% accuracy
for H$_o$ was designated as one of HST's three ``Key Projects''. (The
other two were a study of the intergalactic medium using quasar
absorption lines, and a ``medium-deep'' survey of galaxies.)

This review is organized as follows: We first give a brief overview of
the cosmological context for measurements of the Hubble constant.  We
discuss in some detail methods for measuring distances to galaxies,
specifically Cepheids, the tip of the red giant branch (TRGB), masers,
the Tully-Fisher relation and Type Ia supernovae (SNe~Ia).  We then
turn to a discussion of H$_o$, its systematic uncertainties, other
methods for measuring H$_o$, and future measurements of the Hubble
constant.  Our goal is to describe the recent developments that have
resulted in a convergence to better than 10\% accuracy in measurements
of the Hubble constant, and to outline how future data can improve
this accuracy. For wide-ranging previous reviews of this subject,
readers are referred to those of Hodge (1982), Huchra (1992), Jacoby
et al. (1992), van den Bergh (1992), Jackson (2007), and Tammann,
Sandage \& Reindl (2008).  An extensive monograph by Rowan-Robinson
(1985) details the history of the subject as it stood twenty-five
years ago.

\section{EXPANSION OF THE UNIVERSE: THE COSMOLOGICAL CONTEXT}
\label{sec:cosmology}

Excellent introductions to the subject of cosmology can be found in
Kolb \& Turner (1990) and Dodelson (2003). We give  a brief
description here to provide  the basis for the nomenclature used
throughout this review. The expansion of a homogeneous and isotropic
universe can be described by a Friedmann-Lemaitre-Robertson-Walker
(FLRW) cosmology, which is characterized by parameters that describe
the expansion, the global geometry, and the general composition of the
universe.  These parameters are all related via the Friedmann
equation, derived from the Einstein general relativity field
equations:
\begin{eqnarray}
H^2(t) =  \left( {\dot a\over a} \right)^2 & = & { 8\pi G \over
      3} \sum_i \rho_i(t) - {k\over a^2}
\end{eqnarray}
where $H(t)$ is the expansion rate, G is the Newtonian gravitational
constant, $a(t)$ is the cosmic scale factor characterizing the
relative size of the universe at time $t$ to the present scale,
$\rho_i(t)$ are the individual components of the matter-energy
density, and $k$ (with values of +1, 0, or -1) describes the global
geometry of the universe.  The density $\rho_{i}$ characterizes the
matter-energy composition of the universe: the sum of the densities of
baryons, cold dark matter, and hot dark matter, and the contribution
from dark energy.  Dividing by H$^2$, we may rewrite the Friedmann
equation as $\Omega_{total}$ - 1 = $\Omega_k$ = k/(a$^2$H$^2$).  For
the case of a spatially flat universe ($k = 0$), $\Omega_{total}$ = 1.

In a matter-dominated universe, the expansion velocity of the Universe
slows down over time owing to the attractive force of
gravity. However, a decade ago two independent groups (Perlmutter et
al. 1999; Riess et al. 1998) found that supernovae at z$\sim$0.5
appear to be about 10\% fainter than those observed locally,
consistent instead with models in which the expansion velocity is
increasing; i.e., a universe that is accelerating in its
expansion. Combined with independent estimates of the matter density,
these results are consistent with a universe in which one third of the
overall density is in the form of matter (ordinary plus dark), and two
thirds is in a form having a large, negative pressure, termed dark
energy.  In this current standard model the expansion rate of the
Universe is given by
\begin{eqnarray}
H^2(z)/H_\circ^2 = \Omega_{matter}(1+z)^3+\Omega_{DE}(1 + z)^{3(1+w)}
\end{eqnarray}
\noindent
where $\Omega_{matter}$ and $\Omega_{DE}$ refer to the densities of
(ordinary, cold and hot dark) matter and dark energy, respectively,
and $w = p / \rho$ is the equation of state of the dark energy, the
ratio of pressure to energy density. Recent observations by the
Wilkinson Microwave Anisotropy Probe (WMAP), based on entirely
independent physics, give results  consistent with the supernova data
(Komatsu et al. 2009; Dunkley et al. 2009).  Under the assumption of a
flat universe, the current observations of distant supernovae and
measurements by the WMAP satellite are consistent with  a cosmological
model where $\Omega_{matter}$ = 0.3, $\Omega_{vacuum}$ = 0.7, and $w =
-1$.  The observations are inconsistent with cosmological models
without dark energy.

Another critical equation from general relativity involving the second
derivative of the scale factor is:
\begin{eqnarray}
\ddot a / a = -4\pi~G \sum_i (\rho_i + 3p_i)
\end{eqnarray}
where the sum is over the different contributions to the mass-energy
density of the Universe.  According to this equation, both energy and
pressure govern the dynamics of the Universe, unlike the case of
Newtonian gravity where there is no pressure term. It also allows the
possibility of negative pressure, resulting in an effective repulsive
gravity, consistent with the observations of the acceleration.

Any component of the mass-energy density can be parameterized by its
ratio of pressure to energy density, $w$.  For ordinary matter $w$ =
0, for radiation $w$ = 1/3, and for the cosmological constant $w$ =
-1. The effect on $\ddot a / a$ of an individual component is -4$\pi G
\rho_i(1+3w_i)$.  If $w < -1/3$ that component will drive an
acceleration (positive $\ddot a$) of the Universe. The time evolution
of the equation of state is unknown; a convenient, simple
parameterization is $w(a) = w_o + (1-a)w_a$, where $w_o$ characterizes
the current value of $w$ and $w_a$ its derivative.

\section{MEASUREMENT OF DISTANCES}

In making measurements of extragalactic distances, objects are being
observed at a time when the scale factor of the Universe, $a$, was
smaller, and the age of the Universe, $t$, was younger than at
present. Measuring the cosmic expansion generally involves use of one
of two types of cosmological distances: the luminosity distance,
\begin{eqnarray}
d_\mathrm{L} = \sqrt {L \over {4\pi F} }
\end{eqnarray}
which relates the observed flux (integrated over all frequencies), $F
$,  of an object to its intrinsic
luminosity, $L$,  emitted in its rest frame;  and the angular  
diameter distance,
\begin{eqnarray}
d_\mathrm{A} = {D \over \theta}
\end{eqnarray}
which relates the apparent angular size of an object in radians,
$\theta$, to its proper size, $D$. The luminosity and angular
diameter distances are related by:
\begin{eqnarray}
   d_\mathrm{L}=(1+z)^2d_\mathrm{A}.
\end{eqnarray}
The distance modulus, $\mu$,  is related to the luminosity distance as
follows:
\begin{eqnarray}
   \mu \equiv m - M = 5 ~log ~d_\mathrm{L} - 5
\end{eqnarray}
where m and M  are the apparent  and absolute magnitudes of the objects,
respectively, and d$_\mathrm{L}$ is in units of parsecs.

The requirements for measuring an accurate value of H$_o$ are simple
to list in principle, but are extremely difficult to meet in
practice. The measurement of radial velocities from the displacement
of spectral lines is straightforward; the challenge is to measure
accurate distances.  Distance measurements must be obtained far enough
away to probe the smooth Hubble expansion (i.e., where the random
velocities induced by gravitational interactions with neighboring
galaxies are small relative to the Hubble velocity), and nearby enough
to calibrate the absolute, not simply the relative distance scale.
The objects under study also need to be sufficiently abundant that
their statistical uncertainties do not dominate the error budget.
Ideally the method has a solid physical underpinning, and is
established to have high internal accuracy, amenable to empirical
tests for systematic errors.

We discuss in detail here three high-precision methods for measuring
distances to nearby galaxies: Cepheids, the tip of the red giant
branch (TRGB) method, and maser galaxies.  For more distant galaxies,
we will additionally discuss three methods in detail: the Tully-Fisher
(TF) relation for spiral galaxies, the surface brightness fluctuation
(SBF) method and the maximum luminosities of Type Ia supernovae
(SNe~Ia).  Although maser distances have so far only been published
for two galaxies (NGC~4258 and UGC~3789), this method has considerable
potential, perhaps even at distances that probe the Hubble flow
directly.

Over the preceding decades a large number of other ``distance
indicators'' have been explored and applied with varying degrees of
success, often over relatively restricted ranges of distance. Main
sequence fitting, red giant ``clump'' stars, RR Lyrae stars, the level
of the horizontal branch, Mira variables, novae and planetary nebula
luminosity functions (PNLF), globular cluster luminosity functions
(GCLF), as well as red and blue supergiant stars all fall into this
class. Some, like the RR Lyrae stars, have provided crucial tests for
consistency of zero points but cannot themselves reach very far beyond
the Local Group because of their relatively faint intrinsic
luminosities.  The reader is referred to recent papers on the SN II
distance scale (Dessart \& Hiller 2005); PNLF (Ciardullo et al. 2002);
and Fundamental Plane (FP; Blakeslee et al. 2002) and references
therein.

Our goal here is not to provide an exhaustive review of all distance
determination methods, but rather to focus on a few methods with
demonstrably low dispersion, some currently understood physical basis,
and with sufficient overlap with other methods to test quantitatively
the accuracy of the calibration, and level of systematic errors for
the determination of H$_o$. Before turning to a discussion of methods
for measuring distances, we discuss the general issue of interstellar
extinction.

\subsubsection{Interstellar Extinction}

Interstellar extinction will systematically decrease a star or
galaxy's apparent luminosity.  Thus, if extinction is not corrected
for it will result in a derivation of distance that is systematically
too large. Dust may be present along the line of sight either within
our Milky Way galaxy and/or along the same extended line of sight
within the galaxy under study.

Two main observational paths to correct for interstellar extinction
have been pursued: (1) make observations in at least two wavelength
bands and, using the fact that extinction is a known function of
wavelength to solve explicitly for the distance and
color-excess/extinction effects, or (2) observe at the longest
wavelengths practical so as to minimize implicitly the extinction
effects.  The former assumes prior knowledge of the interstellar
extinction law and carries the implicit assumption that the extinction
law is universal. The latter path is conceptually more robust, given
that it simply makes use of the (empirically-established) fact that
extinction decreases with increasing wavelength. However, working at
longer and longer wavelengths has been technically more challenging so
this path has taken longer in coming to fruition.

From studies of Galactic O \& B stars it is well-established that
interstellar extinction is wavelength dependent, and from the optical
to mid-infrared wavelengths it is a generally decreasing function of
increasing wavelength (see Cardelli, Clayton \& Mathis 1996; Draine et
al. 2003 and references therein for empirical and theoretical
considerations).  Limited studies of stars in external galaxies
(primarily the LMC and SMC) support this view, with major departures
being confined to the ultraviolet region of the spectrum (particularly
near 2200\AA). Both for practical reasons (that is, detector
sensitivity) and because of the nature of interstellar extinction, the
majority of distance-scale applications have avoided the ultraviolet,
so the most blatant changes in the interstellar extinction curve have
been of little practical concern.  At another extreme, the
universality of the longer-wavelength (optical through infrared)
portion of the extinction curve appears to break down in regions of
intense star formation and extremely high optical depths within the
Milky Way. However, the general (diffuse) interstellar extinction
curve, as parameterized by ratios of total-to-selective absorption,
such as $R_V$ = $A_V/E(B-V)$, appears to be much more stable from
region to region. Fortunately Cepheids, TRGB stars, and supernovae are
generally not found deeply embedded in very high optical-depth dust,
but are sufficiently displaced from their original sites of star
formation that they are dimmed mostly by the general, diffuse
interstellar extinction.

\subsection{Cepheid Distance Scale}

Since the discovery of the Leavitt Law (Leavitt, 1908; Leavitt \&
Pickering 1912) and its use by Hubble to measure the distances to the
Local Group galaxies, NGC 6822 (Hubble 1925), M33 (Hubble 1926) and
M31 (Hubble 1929b), Cepheid variables have remained a widely
applicable and powerful method for measuring distances to nearby
galaxies.  In 2009, the American Astronomical Society Council passed a
resolution recognizing the 100th anniversary of Henrietta Leavitt's
first presentation of the Cepheid Period-Luminosity relation (Leavitt
1908). The Council noted that it was pleased to learn of a resolution
adopted by the organizers of the Leavitt symposium, held in November,
2008 at the Harvard-Smithsonian Center for Astrophysics, Cambridge,
MA. There, it was suggested that the Cepheid Period-Luminosity
relation be referred to as the Leavitt Law in recognition of Leavitt's
fundamental discovery, and we do so here.

Cepheids are observed to pulsate with periods ranging from
2 to over 100 days, and their intrinsic brightnesses correlate with
those periods, ranging from -2 $<$ M$_V$ $<$ -6~mag. The ease of
discovery and identification of these bright, variable supergiants
make them powerful distance indicators.  Detailed reviews of the
Cepheid distance scale and its calibration can be found in Madore \&
Freedman (1991), Sandage \& Tammann (2006), Fouque et al. (2007) and
Barnes (2009).  A review of the history of the subject is given
by Fernie (1969).

There are many steps that must be taken in applying Cepheids to the
extragalactic distance scale.  The Cepheids must be identified against
the background of fainter, resolved and unresolved stars that
contribute to the surrounding light of the host galaxy. Overcoming
crowding and confusion is the key to the successful discovery,
measurement and use of Cepheids in galaxies beyond the Local Group.
From the ground, atmospheric turbulence degrades the image resolution,
decreasing the contrast of point sources against the background. In
space the resolution limit is set by the aperture of the telescope and
the operating wavelengths of the detectors. HST gives a factor of ten
increased resolution over most groundbased telescopes of comparable
and larger aperture.

As higher precision data have been accumulated for Cepheids in greater
numbers and in different physical environments, it has become possible
to search for and investigate a variety of lower level, but
increasingly important, systematics affecting the Leavitt Law. Below
we briefly discuss these complicating effects (reddening and
metallicity, in specific) and their uncertainties, and quantify their
impact on the extragalactic distance scale. We then elaborate on
methods for correcting for and/or mitigating their impact on distance
determinations. But first we give an overview of the physical basis
for the Cepheid period-luminosity relation in general terms.

\subsubsection{Underlying Physics}

The basic physics connecting the luminosity and color of a Cepheid to
its period is well understood. Using the Stephan-Boltzmann  law
\begin{eqnarray}
L = 4\pi R^2 \sigma T_e^4
\end{eqnarray}
the bolometric luminosities, $L$, of all stars, including Cepheids,
can be derived. Expressed in magnitudes, the Stefan-Boltzmann  Law  
becomes
\begin{eqnarray}
M_{BOL} = -5~ log ~R - 10 ~log ~T_e + C .
\end{eqnarray}
Hydrostatic equilibrium can be achieved for long periods of time along
the core-helium-burning main sequence.  As a result stars are
constrained to reside there most of the time, thereby bounding the
permitted values of the independent radius and temperature variables
for stars in the M$_{BOL}$ - $log T_e$ plane.

If~ $log ~T_e$ is mapped into an observable intrinsic color (i.e.,
$(B-V)_o$ or $(V-I)_o$) and radius is mapped into an observable period
(through a period-mean-density relation), the period-luminosity-color
(PLC) relation for Cepheids can be determined (e.g., Sandage 1958;
Sandage \& Gratton 1963; and Sandage \& Tammann 1968).  In its
linearized form for pulsating variables, the Stefan-Boltzmann law
takes on the following form of the PLC: $M_V ~=~ \alpha ~logP ~+~
\beta (B-V)_o ~+~ \gamma$.

Cepheid pulsation occurs because of the changing atmospheric opacity
with temperature in the doubly-ionized helium zone. This zone acts
like a heat engine and valve mechanism. During the portion of the
cycle when the ionization layer is opaque to radiation that layer
traps energy resulting in an increase in its internal pressure. This
added pressure acts to elevate the layers of gas above it resulting
in the observed radial expansion. As the star expands it does work
against gravity and the gas cools. As it does so its temperature falls
back to a point where the doubly-ionized helium layer recombines and
becomes transparent again, thereby allowing more radiation to
pass. Without that added source of heating the local pressure drops,
the expansion stops, the star recollapses, and the cycle
repeats. The alternate trapping and releasing of energy in the helium
ionization layer ultimately gives rise to the periodic change in
radius, temperature and luminosity seen at the surface. Not all stars
are unstable to this mechanism. The cool (red) edge of the Cepheid
instability strip is thought to be controlled by the onset of
convection, which then prevents the helium ionization zone from
driving the pulsation.  For hotter temperatures, the helium ionization
zone is located too far out in the atmosphere for significant
pulsations to occur. Further details can be found in the classic
stellar pulsation text book by Cox (1980).

Cepheids have been intensively modeled numerically, with increasingly
sophisticated hydrodynamical codes (for a recent review see Buchler
2009). While continuing progress is being made, the challenges remain
formidable in following a dynamical atmosphere, and in modeling
convection with a time-dependent mixing length approximation. In
general, observational and theoretical period-luminosity-color
relations are in reasonable agreement (e.g., Caputo 2008). However, as
discussed in \S \ref{sec:metallicity}, subtle effects (for example the
effect of metallicity on Cepheid luminosities and colors) remain
difficult to predict from first principles.

\subsubsection{Cepheids and Interstellar Extinction}

If one adopts a mean extinction law and applies it universally to all
Cepheids, regardless of their parent galaxy's metallicity, then one
can use the observed colors and magnitudes of the Cepheids to correct
for the total line-of-sight extinction. If, for example, observations
are made at $V$ and $I$ wavelengths (as is commonly done with HST),
and the ratio of total-to-selective absorption $R_{VI} = A_V/E(V-I)$
is adopted {\it a priori} ({\it e.g.,} Cardelli, Clayton \& Mathis
1989), then one can form from the observed colors and magnitudes an
extinction-free, Wesenheit magnitude, W, (Madore 1982), defined by
\begin{eqnarray}
W = V - R_{VI} \times (V-I)
\end{eqnarray}
\noindent
as well as an intrinsic Wesenheit magnitude, W$_o$
\begin{eqnarray}
W_o  = V_o  - R_{VI} \times (V-I)_o.
\end{eqnarray}
\noindent
By construction
\begin{eqnarray}
W = V_o + A_V - R_{VI} \times (V-I)_o  - R_{VI} \times E(V-I) \\
         = V_o - R_{VI}  (V-I)_o  + A_V - R_{VI} \times E(V-I)
\end{eqnarray}
\noindent
where V = $V_o$ + $A_V$ and (V-I) = $(V-I)_o$ + E(V-I), and $A_V =
R_{VI} \times E(V-I)$, thereby reducing the last two terms to zero,
leaving $W = V_o - R_{VI} \times (V-I)_o$ which is equivalent to the
definition of $W_o$.

The numerical value of W as constructed from observed data points is
numerically identical to the intrinsic (unreddened) value of the
Wesenheit function, $W_o$.  Thus, W, for any given star, is dimmed
only by distance and (by its definition) it is unaffected by
extinction, again only to the degree that R is known and is
universal. W can be formed for any combination of
optical/near-infrared bandpasses.

\subsubsection{Metallicity}
\label{sec:metallicity}

The atmospheres of stars like Cepheids, having effective temperatures
typical of G and K supergiants, are affected by changes in the
atmospheric metal abundance. There are additionally changes in the
overall stellar structure (the mass-radius relation) due to changes in
chemical composition.  Thus, it is expected that the colors and
magnitudes of Cepheids, and their corresponding PL relations, should
be a function of metallicity.  However, predicting the magnitude (and
even simply the sign of the effect) at either optical or even longer
wavelengths, has proven to be challenging: different theoretical
studies have led to a range of conclusions.  We review below the
empirical evidence. For a comparison with recent theoretical studies
we refer the interested readers to papers by Sandage, Bell \& Tripicco
(1999), Bono et al. (2008), Caputo (2008) and Romaniello et al. (2008,
2009).

Two tests of the metallicity sensitivity of the Cepheid PL relation
have been proposed. The first test uses measured radial metallicity
gradients within individual galaxies to provide a differential test in
which  observed changes in the Cepheid zero point with radius are
ascribed to changes in metallicity. This test assumes that the
Cepheids and the HII regions (which calibrate the measured [O/H]
abundances) share the same metallicity at a given radius, and that
other factors are not contributing to a zero-point shift, such as
radially dependent crowding or changes of the extinction law with
radius, etc. The second test uses the difference between Cepheid and
TRGB distances for galaxies with both measurements and seeks a
correlation of these differences as a function of the Cepheid (i.e.,
HII region) metallicity.

The first test, leveraging metallicity gradients in individual
galaxies, has been undertaken for M31 (Freedman \& Madore 1990), M101
(Kennicutt et al. 1998), NGC~4258 (Macri et al. 2006) and M33
(Scowcroft et al. 2009).  The second test, comparing TRGB and Cepheid
distances, was first made by Lee, Freedman \& Madore (1993).  Udalski
et al. (2001) used a newly observed sample of Cepheids in IC~1613 in
comparison to a TRGB distance to that same galaxy, and concluded that,
in comparison with the SMC, LMC and NGC~6822 there was no metallicity
effect over a factor of two in metallicity at low mean metallicity.
An extensive cross comparison of Cepheid and TRGB distances including
high-metallicity systems is well summarized by Sakai et al. (2004).
Individual datasets and metallicity calibrations are still being
debated, but the general concensus is that for the reddening-free
W(VI) calibration of the Cepheid distance scale there is a metallicity
dependence that, once corrected for, increases the distance moduli of
higher metallicity Cepheids if their distances are first determined
using a lower metallicity (e.g., LMC) PL calibration. However, in a
different approach, Romaniello et al (2008) have obtained direct
spectroscopic [Fe/H] abundances for a sample of Galactic, LMC and SMC
Cepheids. 
They compare the Leavitt Law for samples of stars
with different mean metallicities and find a dependence of the V-band
residuals with [Fe/H] abundance that is in the opposite sense to these
previous determinations. Clearly the effect of metallicity on the
observed properties of Cepheids is still an active and on-going area
of research.

A remaining uncertainty at the end of the H$_o$ Key Project 
(described
further in \S \ref{sec:HSTKP}
 was due to the fact that the majority of
Key Project galaxies have metallicities more comparable to the Milky
Way than to the LMC, which was used for the calibration. Below, in \S
\ref{sec:plcalibration} we ameliorate this systematic error by
adopting a Galactic calibration provided by new trigonometric
parallaxes of Milky Way Cepheids, not available at the time of the Key
Project.  By renormalizing to a high-metallicity (Galactic)
calibration for the Cepheids, metallicity effects are no longer a
major systematic, but rather a random error, whose size will decrease
with time as the sample size increases.
Based on the Cepheid
metallicity calibration of Sakai et al (2004) (with adopted LMC and
Solar values for 12 + log (O/H) of 8.50 and 8.70, respectively; and a
metallicity slope of 0.25~mag/dex), we estimate the metallicity
correction in transforming from an LMC to a Galactic-based Cepheid
zero point to be 0.25 x 0.2 = 0.05 mag, with a residual scatter of
about $\pm$ 0.07~mag.

\subsubsection{Galactic Cepheids with Trigonometic Parallaxes}
\label{sec:plcalibration}

An accurate trigonometric parallax calibration for Galactic Cepheids
has been long sought, but very difficult to achieve in practice. All
known classical (Galactic) Cepheids are more than 250~pc away:
therefore for direct distance estimates good to 10\%, parallax
accuracies of $\pm$0.2~ milliarcsec are required, necessitating space
observations. The Hipparchos satellite reported parallaxes for 200 of
the nearest Cepheids, but (with the exception of Polaris) even the
best of these were of very low signal-to-noise (Feast \& Catchpole
1997).

Recent progress has come with the use of the Fine Guidance Sensor on
HST (Benedict et al. 2007), whereby parallaxes, in many cases accurate
to better than $\pm$10\% for individual stars were obtained for 10
Cepheids, spanning the period range 3.7 to 35.6 days. We list the
distance moduli, errors, and distances for these Cepheids in Table
1. These nearby Cepheids span a range of distances from about 300 to
560 pc. 

The calibration of the Leavitt relation based on these ten stars leads
to an error on their mean of $\pm$3\% (or $\pm$0.06~mag), which we
adopt here as the systematic error on the distance to the LMC
discussed below, and the Cepheid zero point, in general. In what
follows, we adopt the zero-point based on the Galactic calibration,
but retain the slope based on the LMC, since the sample size is still
much larger and statistically better defined. Improvement of this
calibration (both slope and zero point) awaits a larger sample of
(long-period) Cepheids from GAIA. We have adopted a zero-point
calibration based both on these HST data, as well as a calibration
based on the maser galaxy, NGC 4258 (\S \ref{sec:masers}) and present
a revised value of H$_o$ in \S \ref{sec:HSTKP}.

A significant systematic at this time is the calibration zero point.
Its value depends on only ten stars, each of which have 
uncertainties in their distances that are individually at the 10\%
level. Given the small sample size of the Galactic calibrators, the
error on their mean can be no better than 3\% (or $\pm$0.06~mag),
which we adopt here as the newly revised systematic error on the
distance to the LMC, and on the Cepheid zero point in general. In what
follows, we adopt the zero point based on the Galactic calibration,
but retain the slope based on the LMC, because the sample size is
still much larger and therefor statistically better defined. There has recently
been discussion in the literature about possible variations in the
slope of the Leavitt Law occurring around 10 days (see Ngeow, Kanbur
\& Nanthakumar (2008) and references therein); however, Riess et
al. (2009a) and Madore \& Freedman (2009) both find that when using W,
the differences are not statistically significant.  Improvement of
this calibration (both in the slope and zero point) awaits a larger
sample of (long-period) Cepheids from the Global Astrometric
Interforometer for Astrophysics satellite (GAIA). 

\subsubsection{The Distance to the Large Magellanic Cloud}
\label{sec:LMC}

Because of the abundance of known Cepheids in the Large Magellanic
Cloud this galaxy has historically played a central role in the
calibration of the Cepheid extragalactic distance scale. Several
thousand Cepheids have been identified and cataloged in the LMC
(Leavitt 1908; Alcock et al. 2000; Soszynski et al. 2008), all at
essentially the same distance.  Specifically, the slope of the Leavitt
Law is both statistically and systematically better determined in the
LMC than it is for Cepheids in our own Galaxy. This is especially true
for the long-period end of the calibration where the extragalactic
samples in general are far better populated than the more restricted
Milky Way subset available in close proximity to the Sun. In Figure 2
we show the range of values of LMC distance moduli based on
non-Cepheid moduli, published up to 2008.  The median value of the
non-Cepheid distance moduli is 18.44$\pm$0.16 mag.

Based on the new results for direct geometric parallaxes to Galactic
Cepheids (Benedict et al. 2007) discussed in \S
\ref{sec:plcalibration}, we calibrate the sample of LMC Cepheids used
as fiducial for the HST Key Project. The new Galactic parallaxes now
allow a zero point to be obtained for the Leavitt Law. In Figure 3, we
show BVIJHK Leavitt Laws for the Galaxy and LMC calibrated with the
new parallaxes.  As can be seen, the slope of the Leavitt Law
increases with increasing wavelength, with a corresponding decrease in
dispersion. In the past, because of the uncertainty in the Galactic
Cepheid calibration, a distance modulus to the LMC and the mean
Cepheid extinction were obtained using a combination of several
independent methods.  Multi-wavelength Leavitt Laws were then used to
obtain differential extragalactic distances and reddenings for
galaxies beyond the LMC. We can show here for the first time the
multiwavelength solution for the distance to the LMC itself based on
the apparent BVIJHK Cepheid distance moduli, fit to a Cardelli et
al. (1989) extinction curve, and adopting a Galactic calibration for
the zero point, and  the slope from the LMC data.  The LMC
apparent moduli, scaled to the Galactic calibration are shown as a
function of inverse wavelength in Figure 4. The data are well fit by a
Galactic extinction law having a scale factor corresponding to E(B-V)
= 0.10~mag, and an intercept at 1/$\lambda$ = 0.00, corresponding to a
true modulus of $\mu(LMC)_o$ = 18.40~$\pm$0.01 mag.

The composite (Galactic + LMC) VI Wesenheit function is shown in
Figure 5.  The correspondence between the two independent Cepheid
samples is good, and the dispersion in W remains very small. The
Wesenheit function uses fewer wavelengths, but it employs the two
bandpasses directly associated with the HST Key Project and most
extragalactic Cepheid distances, and so we adopt it here.

The W(V,VI) Wesenheit function gives a minimized fit between the
Galactic and the LMC Cepheids corresponding to a true distance modulus
of $\mu(LMC)_o$ = 18.44~$\pm$0.03~mag.  Correcting for metallicity
(see \S \ref{sec:metallicity}) would decrease this to 18.39~mag.
Because of the large numbers of Cepheids involved over numerous
wavelengths, the statistical errors on this value are small; and once
again systematic errors dominate the error budget. As discussed in \S
\ref{sec:plcalibration}, we adopt a newly revised systematic error on
the distance to the LMC, of 3\% (or $\pm$0.06~mag).

As noted above, the main drawback to using the LMC as the fundamental
calibrator of the Leavitt Law is the fact that the LMC Cepheids are of
lower metallicity than  many of the more distant spiral galaxies
useful for measuring the Hubble constant.  This systematic is largely
eliminated by adopting the higher-metallicity Galactic calibration as
discussed in \S \ref{sec:metallicity}, or the NGC 4258 calibration
discussed in \S \ref{sec:masers}.

\subsection{Tip of the Red Giant Branch (TRGB) Method}

As discussed briefly in \S \ref{sec:metallicity} a completely
independent method for determining distances to nearby galaxies that
has comparable precision to Cepheids is the tip of the red giant
branch (TRGB).  The TRGB method uses the theoretically well-understood
and observationally well-defined discontinuity in the luminosity
function of stars evolving up the red giant branch in old, metal-poor
stellar populations. This feature has been calibrated using Galactic
globular clusters, and because of its simplicity and straightforward
application it has been widely used to determine distances to nearby
galaxies. A recent and excellent review of the topic is given by
Rizzi et al. (2007) and Bellazzini (2008).

Using the brightest stars in globular clusters to estimate distances
has a long history (ultimately dating back to Shapley 1930 and later
discussed again by Baade 1944).  The method gained widespread
application in a modern context in two papers, one by Da Costa \&
Armandroff (1990) (for Galactic globular clusters), and the other by
Lee, Freedman \& Madore (1993) (where the use of a quantitative
digital filter to measure the tip location was first introduced in a
extragalactic context). 

Approximately 250 galaxies have had their distances measured by the
TRGB method, compared to a total of 57 galaxies with Cepheid
distances. (A comprehensive compilation of direct distance
determinations is available at the following web site:
http://nedwww.ipac.caltech.edu/level5/NED1D/ned1d.html). In practice,
the TRGB method is observationally a much more efficient technique,
since, unlike for Cepheid variables, there is no need to follow them
through a variable light cycle: a single-epoch observation, made at
two wavelengths (to provide color information) is sufficient. A recent
example of applying the TRGB technique to the maser galaxy, NGC~4258,
is shown in Figure 6.

\subsubsection{Theory}
The evolution of a post-main-sequence low-mass star up the red giant
branch is one of the best-understood phases of stellar evolution
(e.g., Iben \& Renzini 1983). For the stars of interest in the context
of the TRGB, a helium core forms at the center, supported by electron
degeneracy pressure. Surrounding the core, and providing the entire
luminosity of the star is a hydrogen-burning shell. The ``helium ash''
from the shell increases the mass of the core systematically with
time. In analogy with the white dwarf equation of state and the
consequent scaling relations that interrelate core mass, M$_c$, and
core radius, R$_c$, for degenerate electron support, the core (=
shell) temperature, T$_c$, and the resulting shell luminosity are
simple functions of M$_c$ and R$_c$ alone: $ T_c \sim M_c / Rc $ and $
L_c \sim M_c^7 / R_c^5.$ As a result, the core mass increases, the
radius simultaneously shrinks and the luminosity increases due to both
effects.  The star ascends the red giant branch with increasing
luminosity and higher core temperatures.  When $T_c$ exceeds a
physically well-defined temperature, helium ignites throughout the
core.  The helium core ignition does not make the star brighter, but
rather it eliminates the shell source by explosively heating and
thereby lifting the electron degeneracy within the core. This dramatic
change in the equation of state is such that the core flash (which
generates the equivalent instantaneous luminosity of an entire galaxy)
is internally quenched in a matter of seconds, inflating the core and
settling down to a lower-luminosity, helium core-burning main
sequence.  The transition from the red giant to the horizontal branch
occurs rapidly (within a few million years) so that observationally
the TRGB can be treated as a physical discontinuity.  A stellar
evolutionary ``phase change'' marks the TRGB.  The underlying power of
the TRGB is that it is a physically-based and theoretically
well-understood method for determining distance. Nuclear physics
fundamentally controls the stellar luminosity at which the RGB is
truncated, essentially independent of the chemical composition and/or
residual mass of the envelope sitting above the core.

The radiation from stars at the TRGB is redistributed with wavelength
as a function of the metallicity and mass of the envelope.
Empirically it is found that the bolometric corrections are smallest
in the $I$-band, and most recent measurements have been made at this
wavelength.  The small residual metallicity effect on the TRGB
luminosity is well documented, and can be empirically calibrated out
(see Madore, Mager \& Freedman 2009).

\subsubsection{Recent TRGB Results and Calibration of H$_o$}

In the context of measuring the Hubble constant, RGB stars are not as
bright as Cepheids, and therefore cannot be seen as far, but they can
still be seen to significant distances ($\sim$20~Mpc and including
Virgo, e.g., Durrell et al. 2007; Caldwell 2006) and, as we have
seen, they can serve an extremely important function as an independent
test of the Cepheid distance scale and check on systematic effects.

Mould \& Sakai (2008) have used the TRGB as an alternate calibration
to the Cepheid distance scale for the determination of H$_o$.
They use 14 galaxies for which TRGB distances can be measured
to calibrate the Tully-Fisher relation, and determine a value of $H_
\circ$ = 73 $\pm$ 5 (statistical only) km s$^{-1}$ Mpc$^{-1}$, a value
about 10\% higher than found earlier by Sakai et al. (2000) based on a
Cepheid calibration of 23 spiral galaxies with Tully-Fisher
measurements. In subsequent papers they calibrate the SBF method
(Mould \& Sakai 2009a) and then go on to check the calibration of the
FP for early-type galaxies and the luminosity scale of Type Ia
supernovae (Mould \& Sakai 2009b). They conclude that the TRGB and
Cepheid distances scales are all consistent using SBF, FP, SNe Ia and
the TF relation.

\subsection{Maser Galaxies}
\label{sec:masers}

$H_2O$ mega-masers have recently been demonstrated to be a powerful
new geometric tool for accurately measuring extragalactic
distances. An extensive review of both the physical nature and the
application of mega-masers to the extragalactic distance scale can be
found in Lo (2005). The technique utilizes the mapping of 22.2 GHz
water maser sources in the accretion disks of massive black holes
located in spiral galaxies with active galactic nuclei, through
modeling of a rotating disk ideally in pure Keplerian motion.  In the
simplest version of the technique, a rotation curve is measured along
the major axis of the disk; proper motions are measured on the near
side of the disk minor axis, and a comparison of the angular
velocities in the latter measurement with the absolute velocities in
km s$^{-1}$ in the former measurements yields the distance.

The method requires a sample of accretion disks that are relatively
edge on (so that a rotation curve can be obtained from radial-velocity
measurements) and a heating source such as x-rays or shocks to produce
maser emission. The basic assumption is that the maser emission arises
from trace amounts of water vapor ($<$10$^{-5}$ in number density) in
very small density enhancements in the accretion disk and that they
act as perfect dynamical test particles. The maser sources appear as
discrete peaks in the spectrum or as unresolved spots in the images
constructed from Very Long Baseline Interferometry
(VLBI). Measurements of the acceleration (a = V$^2$/ r) are obtained
directly by monitoring the change of maser radial velocities over time
from single-dish observations. Proper motions are obtained from
observed changes in angular position in interferometer images. The
approximately Keplerian rotation curve for the disk is modeled,
allowing for warps and radial structure.  The best studied galaxy, NGC
4258, at a distance of about 7 Mpc, is too close to provide an
independent measurement of the Hubble constant (i.e., free from local
velocity-field perturbations) but it serves as an invaluable
independent check of the Cepheid zero-point calibration.

\subsubsection{A Maser Distance  to NGC~4258}
\label{sec:masergal}

VLBI observations of $H_2O$ maser sources surrounding the active
galactic nucleus of NGC 4258 reveal them to be in a very thin,
differentially rotating, slightly warped disk.  The Keplerian velocity
curve has deviations of less than one percent. The disk has a
rotational velocity in excess of 1,000 km/s at distances on the order
of 0.1 pc from the inferred super-massive (10$^7$M$_\odot$) nuclear
black hole. Detailed analyses of the structure of the accretion disk
as traced by the masers have been published (e.g., Herrnstein et
al. 1999; Humphreys, et al, 2008 and references therein). Over time it
has been possible to measure both proper motions and accelerations of
these sources and thereby allow for the derivation of two independent
distance estimates to this galaxy. The excellent agreement of these
two estimates supports the {\it a priori} adoption of the Keplerian
disk model and gives distances of 7.2 $\pm$ 0.2 and 7.1 $\pm$ 0.2~Mpc,
respectively.

Because of the simplicity of the structure of the maser system in NGC
4258 and its relative strength, NGC 4258 will remain a primary test
bed for studying systematic effects that may influence distance
estimates. Several problems may limit the ultimate accuracy of this
technique, however. For example, because the masers are only
distributed over a small angular part of the accretion disk, it is
difficult to assess the importance of non-circular orbits.  Of
possible concern, eccentric disks of stars have been observed in a
number galactic nuclei where the potential is dominated by the black
hole, as is the case for NGC 4258.  In addition, even if the disk is
circular, it is not a given that the masers along the minor axis are
at the same radii as the masers along the major axis.  The self
gravity of the disk also may need to be investigated and modeled since
the maser distribution suggests the existence of spiral arms
(Humphreys et al., 2008). Finally, radiative transfer effects may
cause non-physical motions in the maser images.  Although the current
agreement of distances using several techniques is comforting, having
only one sole calibrating galaxy for this technique remains a concern,
and further galaxies will be required to ascertain the limiting
uncertainty in this method.

\subsubsection{Other Distance Determinations to NGC~4258}

The first Cepheid distance to NGC~4258 was published by Maoz et al,
(1999) who found a distance of 8.1$\pm$0.4 Mpc, based on an
LMC-calibrated distance modulus of 18.50~mag.  Newman et al. (2001)
found a distance modulus of 29.47 $\pm$ 0.09 (random) $\pm$ 0.15
(systematic) giving a distance of 7.83 $\pm$ 0.3 $\pm$0.5 Mpc.  Macri
et al. (2006) reobserved NGC~4258 in two radially (and chemically)
distinct fields discovering 281 Cepheids at BV and I
wavelengths. Their analysis gives a distance modulus of 29.38 $\pm$
0.04 $\pm$0.05~mag (7.52 $\pm$ 0.16~Mpc), if one adopts $\mu (LMC) =
18.50$~mag.  Several more recent determinations of resolved-star
(Cepheid and TRGB) distance moduli to NGC~4258 are in remarkably
coincident agreement with the maser distance modulus. di Benedetto
(2008) measures a Cepheid distance modulus of 29.28 $\pm$0.03
$\pm$0.03 for NGC~4258, corresponding to a distance of 7.18~Mpc;
Benedict et al. (2007) also find a distance modulus of 29.28 $\pm$ 08
mag ; and Mager, Madore \& Freedman (2008) also find a value of 29.28
$\pm$0.04 $\pm$0.12~mag both from Cepheids and from the TRGB
method. These latter studies are in exact agreement with the current
maser distance.  Higher accuracy has come from larger samples with
higher signal-to-noise data, and improved treatment of metallicity.

An alternative approach to utilizing the maser galaxy in the distance
scale is to adopt the geometric distance to NGC~4258 as foundational,
use it to calibrate the Leavitt Law, and from there determine the
distance to the LMC. Macri et al. (2006) adopted this approach and
conclude that the true distance modulus to the LMC is
18.41$\pm$0.10~mag. This value agrees well with the new Galactic
Cepheid calibration of the LMC Leavitt law, as discussed in \S
\ref{sec:LMC}.

\subsubsection{NGC~4258 and the Calibration of H$_o$}

The distance to NGC~4258 can be used to leapfrog over the LMC
altogether to calibrate the Cepheid PL relation and then secondary
methods.  Macri et al. (2006) and Riess et al. (2009a,b) have adopted
the distance to NGC~4258 as a calibration of the supernova distance
scale, as discussed further in \S \ref{sec:snh0}.

Attempts to measure distances to other megamasers has proved to be
difficult.  About 2000 galaxies have been surveyed for masers and more
than 100 masers discovered to date. The detection rate of about 5\% is
likely due to detection sensitivity and the geometric constraint that
the maser disk be viewed nearly edge on because the maser emission is
expected to be highly beamed in the plane of the disk. About 30 of
these masers have spectral profiles indicative of emission from thin
disks: i.e., masers at the galactic systemic velocity and groups of
masers symmetrically spaced in velocity. About a dozen maser galaxies
are sufficiently strong that they can be imaged with phase-referenced
VLBI techniques. Only about five have been found to have sufficiently
simple structure so that they can be fit to dynamical models and have
their distances determined.  The most promising examples of these
galaxies is UGC 3789, which has a recessional velocity of greater than
3000 km/s, and is being pursued by the Megamaser Cosmology Project
(Reid et al.  2009).

If a significant number of maser galaxies can be found and precisely
observed well into the Hubble flow, this method, can, in principle,
compete with methods such as SNe Ia for measuring H$_\circ$. The
challenge will be to obtain large enough sample sizes of hundreds of
objects, in order to average over large-scale flows. Unfortunately,
this likely will not be accomplished in the upcoming decade. It is
also hoped that nearby objects will be found where this technique can
be applied, in addition to NGC 4258, and strengthen the zero-point
calibration of the extragalactic distance scale. The future for this
technique (beyond 2020) looks promising, given a high- frequency
capability for the  Square Kilometer Array.

\subsection{Surface Brightness Fluctuation (SBF) Method}

For distances to elliptical galaxies and early-type spirals with large
bulge populations the Surface Brightness Fluctuation (SBF) method,
first introduced by Tonry and Schneider (1988), overlaps with and
substantially exceeds the current reach of the TRGB method. Both
methods use properties of the red giant branch luminosity function to
estimate distances. The SBF method quantifies the effect of distance
on an over-all measure of resolution of the Population II red giant
stars, naturally weighted both by their intrinsic luminosities and
relative numbers. What is measured is the pixel-to-pixel variance in
the photon statistics (scaled by the surface brightness) as derived
from an image of a pure population of red giant branch stars. For
fixed surface brightness, the variance in a pixel (of fixed angular
size) is a function of distance, simply because the total number of
discrete sources contributing to any given pixel increases with the
square of the distance. While the TRGB method relies entirely on the
very brightest red giant stars, the SBF method uses a
luminosity-weighted integral over the entire RGB population in order
to define a typical ``fluctuation star'' whose mean magnitude,
$\overline{M_I}$ is assumed to be universal and can therefore be used
to derive distances. For recent discussions of the SBF method, the
reader is referred to Biscardi et al. (2008) and Blakeslee et
al. (2009).

Aside from the removal of obvious sources of contamination such as
foreground stars, dust patches and globular clusters, the SBF method
does require some additional corrections. It is well known that the
slope of the red giant branch in the color-magnitude diagram is a
function of metallicity, and so the magnitude of the fluctuation star
is both expected and empirically found to be a function metallicity. A
(fairly steep) correction for metallicity has been derived and can be
applied using the mean color of the underlying stellar population
$\overline{M_I} = -1.74 +4.5(V-I)_o -1.15$ (Tonry et al. 2002).

A recent and comprehensive review of the application of the SBF method
to determining cosmic distances, and its comparison to the
Fundamental-Plane (FP) method is given in Blakeslee et al (2002).
Over 170 galaxies enter into the comparison; this analysis leads to
the conclusion that $H_o = 72 \pm 4 (random) \pm 11 (systematic)$
km/s/Mpc.

\subsection{Tully-Fisher Relation}

The total luminosity of a spiral galaxy (corrected to face-on
inclination to account for extinction) is strongly correlated with the
galaxy's maximum (corrected to edge-on inclination) rotation
velocity. This relation, calibrated via the Leavitt Law or TRGB,
becomes a powerful means of determining extragalactic distances (Tully
\& Fisher 1977; Aaronson et al. 1986; Pierce \& Tully 1988; Giovanelli
et al. 1997). The Tully-Fisher relation at present is one of the most
widely applied methods for distance measurements, providing distances
to thousands of galaxies both in the general field and in groups and
clusters. The scatter in this relation is wavelength-dependent and
approximately $\pm$0.3-0.4~mag or 15-20\% in distance (Giovanelli et
al. 1997; Sakai et al. 2000; Tully \& Pierce 2000).

In a general sense, the Tully-Fisher relation can be understood in terms
of the virial relation applied to rotationally supported disk
galaxies, under the assumption of a constant mass-to-light ratio
(Aaronson, Mould \& Huchra 1979). However, a detailed self-consistent
physical picture that reproduces the Tully-Fisher relation (e.g.,
Steinmetz \& Navarro 1999), and the role of dark matter in producing
almost universal spiral galaxy rotation curves (McGaugh et al. 2000)
still remain a challenge.

{\it Spitzer} archival data have recently yielded an unexpected and
exciting discovery.  Of the 23 nearby galaxies with HST Cepheid
distances that can be used to independently calibrate the Tully-Fisher
relation, there are eight that currently also have 3.6$\mu$m published
total magnitudes (Dale et al. 2007). In Figure~7 (left three panels)
we show the B, I and H-band TF relations for the entire sample of
currently available calibrating galaxies from Sakai et
al. (2000). Their magnitudes have been corrected for
inclination-induced extinction effects and their line widths have been
corrected to edge-on. The scatter is $\pm$0.43, 0.36 and 0.36~mag for
the B, I and H-band relations, respectively; the outer lines follow
the mean regression at $\pm$2-sigma. If it holds up with further data,
this intrinsic scatter means that to measure a distance good to 5\%,
say, using even the best of these TF relations one would need to find
a grouping of 16 galaxies in order to beat down the intrinsic rms
scatter. In the right panel of Figure~7 we show the mid-IR TF relation
for the eight galaxies with Cepheid distances and published IRAC
observations, measured here at 3.6$\mu$m. The gains are
impressive. With the magnitudes not even corrected for any inclination
effects, the scatter within this sample is found to be only
$\pm$0.12~mag. Each of these galaxies entered the calibration with its
own independently determined Cepheid-calibrated distance. If this
correlation stands the test of time as additional calibrators enter
the regression, using the mid-IR TF relation a single galaxy could
potentially yield a distance good to $\pm$5\%. All TF galaxies, when
observed in the mid-IR, would then individually become precision
probes of large-scale structure, large-scale flows and the Hubble
expansion.

\subsection{Type Ia Supernovae}

One of the most accurate means of measuring cosmological distances out
into the Hubble flow utilizes the peak brightness of Type Ia
supernovae (SNe~Ia).  The potential of supernovae for measuring
distances was clear to early researchers (e.g., Baade, Minkowski,
Zwicky) but it was the Hubble diagram of Kowal (1968) that set the
modern course for this field, followed by decades of work by Sandage,
Tammann and collaborators (e.g., Sandage \& Tammann 1982; Sandage \&
Tammann 1990); see also the review by Branch (1998).  Analysis by
Pskovskii (1984), followed by Phillips (1993), established a
correlation between the magnitude of a SN~Ia at peak brightness and
the rate at which it declines, thus allowing supernova luminosities to
be ``standardized''.  This method currently probes farthest into the
unperturbed Hubble flow, and it possesses very low intrinsic scatter:
in recent studies, the decline-rate corrected SN~Ia Hubble diagram is
found to have a dispersion of $\pm$7-10\% in distance (e.g., Folatelli
et al. 2009, Hicken et al. 2009). A simple lack of Cepheid calibrators
prevented the accurate calibration of Type Ia supernovae for
determination of H$_\circ$ prior to HST. Substantial improvements to
the supernova distance scale have resulted from recent dedicated,
ground-based supernova search and follow-up programs yielding CCD
light curves for nearby supernovae (e.g., Hamuy et al.  1996;
Jha et al. 2006; Contreras et al. 2010). Sandage
and collaborators undertook a major program with HST to find Cepheids
in nearby galaxies that have been host to Type Ia supernovae (Sandage
et al. (1996) Saha et al. 1999), and thereby provided the first
Cepheid zero-point calibration, which has recently been followed up by
Macri et al. (2006) and Riess et al. (2009a,b).

For Hubble constant determinations, the challenge in using SNe~Ia
remains that few galaxies in which SN~Ia events have been observed are
also close enough for Cepheid distances to be measured. Hence, the
calibration of the SN~Ia distance scale is still subject to
small-number statistical uncertainties. At present, the numbers of
galaxies for which there are high-quality Cepheid and SN~Ia
measurements (in most cases made with the same telescopes and
instruments as the Hubble flow set) is limited to six objects (Riess
et al. 2009a).

\subsubsection{Underlying Theory}

SNe~Ia result from the thermonuclear runaway explosions of stars. From
observations alone, the presence of SNe Ia in elliptical galaxies
suggests that they do not come from massive stars.  Many details of
the explosion are not yet well understood, but the generally accepted
view is that of an carbon-oxygen, electron-degenerate,
nearly-Chandrasekhar-mass white dwarf orbiting in a binary system with
a close companion (Whelan \& Iben 1973). As material from the Roche
lobe of the companion is deposited onto the white dwarf, the pressure
and temperature of the core of the white dwarf increases until
explosive burning of carbon and oxygen is triggered.  An alternative
model is that of a ``double degenerate'' system (merger with another
white dwarf). Although on observational grounds, there appear to be
too few white dwarf pairs, this issue has not been conclusively
resolved.  A review of the physical nature of SNe~Ia can be found in
Hillebrandt \& Niemeyer (2000).

A defining characteristic of observed SNe~Ia is the lack of hydrogen
and helium in their spectra. It is presumed that the orbiting
companion is transferring hydrogen- and helium-rich material onto the
white dwarf; however, despite extensive searches this hydrogen or
helium has never been detected, and it remains a mystery as to how
such mass transfer could take place with no visible signature. It is
not yet established whether this is a problem of observational
detection, or whether these elements are lost from the system before
the explosion occurs.

Various models for SN~Ia explosions have been investigated. The most
favored model is one in which a subsonic deflagration flame is
ignited, which subsequently results in a supersonic detonation wave (a
delayed detonation). The actual mechanism that triggers a SN~Ia
explosion is not well understood: successfully initiating a detonation
in a CO white dwarf remains extremely challenging.  In recent years,
modeling in 3D has begun, given indications from spectropolarimetry
that the explosions are not spherically symmetric.  The radiative
transport calculations for exploding white dwarf stars are
complex. However, there is general consensus that the observed
(exponential shape of the) light curves of SN e~Ia are powered by the
radioactive decay of $^{56}$Co to $^{56}$Fe.  The range of observed
supernova peak brightnesses appears to be due to a range in $^{56}$Ni
produced. However, the origin of the peak magnitude - decline rate is
still not well understood.

Despite the lack of a solid theoretical understanding of SNe~Ia,
empirically they remain one of the best-tested, lowest-dispersion, and
highest-precision means of measuring relative distances out into the
smooth Hubble flow.

\subsubsection{Recent Results for SNe Ia and H$_\circ$}
\label{sec:snh0}

The most recent calibration of SNe~Ia has come from Riess et
al. {2009a,b} from a new calibration of six Cepheid distances to
nearby well-observed supernovae using the Advanced Camera for Surveys
(ACS) and the Near-Infrared Camera and Multi-Object Spectrometer
(NICMOS) on HST.  Riess et al. have just completed a program to
discover Cepheids in nearby galaxies known to have been hosts to
relatively recent Type Ia supernovae and then re-observed them in the
near infrared.  In so doing, the number of high-quality calibrators
for the supernova distance scale more than doubled, putting the
calibration for SNe~Ia on a far more secure foundation. The six
Cepheid-calibrated supernovae include SN1981B in NGC 4536, SN 1990N in
NGC 4639, SN 1998aq in NGC 3982, SN 1994ae in NGC 3370, SN 1995al in
NGC 3021 and finally SN 2002fk in NGC 1309.  A comparison of Cepheid
and SNe Ia distances from Riess et al. (2009a) is shown in Figure
8. The supernovae were chosen to meet rather stringent criteria,
requiring, for example that they all were observed with modern
detectors, that they were observed before maximum light, their spectra
were not atypical and that their estimated reddenings were low.  Each
galaxy had between 13 and 26 Cepheids observed at random phases in the
H-band (F160W filter) (and were transformed to mean light using
optical data) using NICMOS onboard HST.  Extinction in the H-band is
down by a factor of five relative to the optical. The program avoids
issues of cross-instrumental calibration by observing with a single
telescope for the calibration galaxy, NGC 4258, out to the SNe Ia
galaxies.  By extending to the near-infrared, these observations of
the newly discovered Cepheids directly address the systematic effects
of metallicity and reddening.

We show in Figure 9, the Hubble diagram for 240 supernovae at z $<$
0.1 from Hicken et al. (2009), which have been calibrated by Riess et
al. (2009a) based on the distance to the maser galaxy, NGC 4258. Riess
et al. quote a value of H$ _o$ = 74.2 $\pm$ 3.6 km s$^{-1}$ Mpc$^{-1}$
combining systematic and statistical errors into one number, a value
in excellent agreement with that from the Key Project (see next
section), which is calibrated using the Galactic Cepheid parallax
sample. At the current time, there is not much need for larger,
low-redshift samples, since the dominant remaining uncertainties are
systematic, rather than statistical.  Recent studies (e.g., Wood- 
Vasey et
al. 2008; Folatelli et al. 2009) confirm that supernovae are better
standard candles at near-infrared (JHK) wavelengths and minimize the
uncertainties due to reddening.

Tammann, Sandage \& Reindl (2008) have undertaken a recent
re-calibration of supernovae, as well as a comparison of the Cepheid,
RR Lyrae and TRGB distance scales. In contrast, they find a value of
$H_\circ = 62.3 \pm 4.0$ ~km/s/Mpc, where the quoted (systematic)
error includes their estimated uncertainties in both the Cepheid and
TRGB calibration zero points.  Their quoted error is dominated by the
systematic uncertainties in the Cepheid zero point and the small
number of supernova calibrators, both of which are estimated by them
to be at the 3-4\% level; however, the H$_o$ values differ by more
than 2-$\sigma$. A discussion of the reason for the differences in
these analyses can be found in Riess et al. (2009a,b): these include
the use of more heavily reddened Galactic Cepheids, the use of less
accurate photographic data and a calibration involving multiple
telescope/instruments for supernovae by Tammann, Sandage \& Reindl.

\section{THE HUBBLE SPACE TELESCOPE (HST) KEY PROJECT}
\label{sec:HSTKP}

We briefly summarize below the results from the HST Key Project, and
provide an updated calibration for these data.  The primary goals of
the HST Key Project were to discover and measure the distances to
nearby galaxies containing Cepheid variables, calibrate a range of
methods for measuring distances beyond the reach of Cepheids to test
for and minimize sources of systematic uncertainty, and ultimately to
measure $H_\circ$ to an accuracy of $\pm$10\%.  HST provided the
opportunity to measure Cepheid distances a factor of 10 more distant
than could be routinely obtained on the ground. It also presented a
practical advantage in that, for the first time, observations could be
scheduled in a way that optimized the discovery of Cepheids with a
range of periods independent of phase of the moon or weather (Madore
\& Freedman 2005).

Cepheid distances to 18 galaxies with distances in the range of 3 to
25 Mpc were measured using WF/PC and (primarily) WFPC2 on
HST. Observations at two wavelengths ($V$- and $I$-band) were made,
chosen to allow corrections for dust. The spacing of observations was
optimized to allow for the discovery of Cepheids with periods in the
range of 10 to 50 days. In addition, 13 additional galaxies with
published Cepheid photometry were analyzed for a total of 31 galaxies.

These Cepheid distances were then used to calibrate the Tully-Fisher
relation for spiral galaxies, the peak brightness of Type Ia SNe, the
D$_n-\sigma$ relation for elliptical galaxies, the Surface Brightness
Fluctuation (SBF) method, and Type II supernovae (Freedman 2001 and
references therein).  These methods allowed a calibration of distances
spanning the range of about 70 Mpc (for SBF) out to about 400 Mpc for
Type Ia SNe.  These results are summarized in Figure 10.  Combining
these results using both Bayesian and frequentist methods yielded a
consistent value of $H_\circ$ = 72 $\pm$ 3 (statistical) $\pm$ 7
(systematic) km s$^{-1}$ Mpc$^{-1}$.

We update this analysis using the new HST-parallax Galactic
calibration of the Cepheid zero point (Benedict et al. 2007), and the
new supernova data from Hicken et al. (2009). We find a similar value
of H$_o$, but with reduced systematic uncertainty, of H$_o$ = 73
$\pm$2 (random) $\pm$4 (systematic)~km~s$^{-1}$~Mpc$^{-1}$.  The
reduced systematic uncertainty, discussed further in \S
\ref{sec:systematics} below, results from having a more robust
zero-point calibration based on the Milky Way galaxy with comparable
metallicity to the spiral galaxies in the HST Key Project
sample. Although, the new parallax calibration results in a shorter
distance to the LMC (which is no longer used here as a calibrator),
the difference in H$_o$ is nearly offset by the fact that no
metallicity correction is needed to offset the difference in
metallicity between the LMC and calibrating galaxies.

\subsection{Systematics on $H_\circ$ at the End of the Key Project and a
Decade Later}
\label{sec:systematics}

A primary goal of the HST Key Project was the explicit propagation of
statistical errors, combined with the detailed enumeration of and
accounting for known and potential systematic errors.  In Table 2 we
recall the systematics error budget given in Freedman et al. (2001).
The purpose of the original tabulation was to clearly identify the
most influential paths to greater accuracy in future efforts to refine
$H_\circ$. Here we now discuss what progress has been made, and what we
can expect in the very near future using primarily space-based
facilities, utilizing instruments operating mainly at mid-infrared and
near-infrared wavelengths.

Identified systematic uncertainties in the HST Key Project
determination of the extragalactic distance scale limited its stated
accuracy to $\pm$10\%. The dominant systematics were: (a) the zero
point of the Cepheid PL relation, which was tied directly to the
(independently adopted) distance to the LMC; (b) the differential
metallicity corrections to the PL zero point in going from the
relatively low-metallicity (LMC) calibration to target galaxies of
different (and often larger) metallicities; (c) reddening corrections
that required adopting a wavelength dependence of the extinction curve
that is assumed to be universal; and (d) zero-point drift, offsets,
and transformation uncertainties between various cameras on HST and on
the ground.  Table 2 compares these uncertainties to what is now being
achieved with HST parallaxes and new HST SNe Ia distances (Table 2,
Column 3 ``Revisions''), and then what is expected to be realized by
extending to a largely space-based near and mid-infrared Cepheid
calibration using the combined power of HST, {\it Spitzer} and
eventually the {\it James Webb Space Telescope} (JWST) and {\it GAIA}.
(Column 4, ``Anticipated'').

In 2001 the uncertainty on the zero point of the Leavitt Law was the
largest on the list of known systematic uncertainties. Recall that the
Key Project zero point was tied directly to an LMC true distance
modulus of 18.50~mag. As we have seen in \S \ref{sec:plcalibration}
improvement to the zero point has come from new HST parallax
measurements of Galactic Cepheids, improved distance measurements to
the LMC from near-infrared photometry, and measurement of a maser
distance to NGC~4258. We adopt a current zero-point uncertainty of
3\%.

We next turn to the issue of metallicity.  As discussed in \S
\ref{sec:metallicity}, in the optical, metallicity corrections remain
controversial. However, by shifting the calibration from the
low-metallicity Cepheids in the LMC to the more representative and
high-metallicity Milky Way (or alternatively to) the NGC 4258
Cepheids, the character of the metallicity uncertainty has changed
from being a systematic to a random uncertainty.  We conservatively
estimate that the systematic component of the uncertainty on the
metallicity calibration should now drop to $\pm$0.05~mag. Including
the recent results from Benedict et al. (2007) and Riess et
al. (2009a,b), our estimate for the current total uncertainty on
H$_\circ$ is $\pm$ 5\%.

In terms of future improvements, as discussed further in \S
\ref{sec:future}, with the Global Astrometric Interferometer for
Astrophysics (GAIA), and possibly the Space Interferometry Mission
(SIM), the sample of Cepheids with high precision trignometric
parallaxes will be increased, and as more long-period Cepheids enter
the calibration both the slope and the zero point of the
high-metallicity Galactic Leavitt Law will be improved.  By
extending both the calibration of the Leavitt Law and its application
to increasingly longer wavelengths the effects of metallicity and the
impact of total line-of-sight reddening, each drop below the
statistical significance threshold.  At mid-infrared wavelengths the
extinction is  a factor of $\sim$20 reduced compared to optical
wavelengths.  And line blanketting in the mid and near infrared is
predicted theoretically to be small compared to the blue portion of
the spectrum.  Direct tests are now being undertaken to establish
whether this is indeed the case and/or calibrate out any residual
impact (\S \ref{sec:futureceph}).

In principle, a value of $H_\circ$ having a well determined systematic
error budget of only 2-3\% is within reach over the next decade.  It
is the goal of the new {\it Carnegie Hubble Program}, described
briefly in \S \ref{sec:futureceph}, based on a mid-infrared
calibration of the extragalactic distance scale using the {\it
Spitzer} satellite, GAIA and JWST.

\section{OTHER METHODS FOR DETERMINING H$_o$}

Although the focus of this review is the determination of H$_o$ and
the extragalactic distance scale, we briefly mention two indirect
techniques that probe great cosmological distances independently:
gravitational lensing and the Sunyaev-Zel'dovich effect. We also
discuss measurements of anisotropies in the cosmic microwave
background, which offer a measurement of H$_0$, in combination with
other data.

\subsection{Gravitational Lens Time Delays and H$_o$}

As first realized by Refsdal (1964), measurements of the differences
in arrival time, coupled with measurements of the angular separation
of strongly lensed images of a time-variable object (such as a quasar
or supernova) can be used to measure H$_o$. The time delay observed
between multiple images is proportional to H$_o^{ -1}$, and is less
dependent on other cosmological parameters such as $\Omega_{matter}$
and $\Omega_\Lambda$.  An extensive review of the physics of lensing
can be found in Blandford \& Narayan (1986); the observational issues
have been summarized nicely by Myers (1999) and Schechter (2005).

Initially, the practical implementation of this method suffered from a
number of difficulties. Time delays have proven difficult to measure
accurately, the amplitude of quasar variability is generally small,
and relatively few lens systems that can be modeled simply and cleanly
have been found. Dust obscuration is an issue at optical wavelengths.
A great challenge of this method is that astronomical lenses are
galaxies whose underlying mass distributions are not known, and a
strong physical degeneracy exists between the mass distribution of the
lens and the value of H$_o$. As emphasized by Gorenstein, Shapiro \&
Falco (1988), the deflections and distortions do not uniquely
determine the mass distribution: a lens may be located in a group(s)
or cluster(s), which will affect the predicted time delays, an effect
termed the mass sheet degeneracy. Measurements of velocity dispersion
as a function of position can be used to constrain the mass
distribution of the lens, but generally only central velocity
dispersion measurements are feasible. An advantage of the method is
that it offers a probe directly at cosmological distances; the
concomittent disadvantage is that the cosmological model must be
assumed in order to determine H$_o$.  Earlier estimates of H$_o$ using
this technique yielded values about 10\% lower (analyzing the same
data), assuming what was then the standard cosmological model with
$\Omega_{matter}$ = 1.0, in comparison to the current standard model
with $\Omega_{matter}$ = 0.3 and $\Omega_\Lambda$ = 0.7.

The precision and accuracy of this technique has continued to improve
over time. A brief survey of results from gravitational lensing over the
past five years can be found in Suyu et al. (2009), with estimates of
H$_o$ in the range 50 to 85 km s$^{-1}$ Mpc$^{-1}$. There is a wide
range in types of modeling and treatment of errors for these different
systems (e.g., assumed isothermal profiles, assumptions about the
density distribution of the environment, and how well the models are
constrained by the data).

A recent extensive analysis of the quadruple lens system B1608+656 has
been carried out by Suyu et al. (2009). This analysis is based on deep
F606W and F814W ACS data, a more accurate measurement of the velocity
dispersion using the Low-Resolution Imaging Spectrometer (LRIS) on
Keck, a more detailed treatment of the lens environment using a
combination of ray tracing through cosmological N-body simulations
(the Millennium Simulation) along with number counts in the field of

B1608+656, in order to help break the mass sheet degeneracy
problem. Adopting the standard cosmological model with $\Omega_
{matter}$ = 0.3, $\Omega_\Lambda$ = 0.7, and w = -1, they find H$_o$ =
71 $\pm$ 3 km s$^{-1}$ Mpc$^{-1}$, a factor of two improvement over
the previous estimate for this lens.

\subsection{The Sunyaev-Zel'dovich (SZ)  Effect and H$_o$}

Sunyaev \& Zel'dovich (1969) described the inverse-Compton scattering
of photons from the cosmic microwave background (CMB) off of hot
electrons in the X-ray gas of rich clusters of galaxies. This
scattering leads to a redistribution of the CMB photons so that a
fraction of the photons move from the Rayleigh-Jeans to the Wien side
of the blackbody spectrum, referred to as the Sunyaev-Zel'dovich (SZ)
effect. The measured effect amounts to about 1 mK.  The Hubble
constant is obtained using the fact that the measured X-ray flux from
a cluster is distance-dependent, whereas the SZ decrement is
essentially independent of distance. Observations of this effect have
improved enormously in recent years, with high signal-to-noise, high
angular resolution, SZ images obtained with ground-based
interferometric arrays and high-resolution X-ray spectra. The theory
of the SZ effect is covered at length by Birkinshaw (1999); a nice
summary of observational techniques and interferometry results is
given in Carlstrom et al. (2002).

The SZ effect is proportional to the first power of the electron
density, n$_e$: $\Delta T_{SZ} \sim \int dl n_e T_e$, where T$_e$ is
the electron temperature, and d$l$ is the path length along the
line-of-sight, related to the angular diameter distance. The X-ray
emission is proportional to the second power of the density: $S_x \sim
\int dl \Lambda n_e^2$, where $\Lambda$ is the cooling function for
the X-ray gas. The angular diameter distance is solved for by
eliminating the electron density (see Carlstrom et al. 2002;
Birkinshaw 1999).

An advantage of this method is that it can be applied at cosmological
distances, well into the Hubble flow. The main uncertainties result
from potential substructure in the gas of the cluster (which has the
effect of reducing H$_o$), projection effects (if the clusters
observed are prolate, the sense of the effect is to increase H$_o$),
the assumption of hydrostatic equilibrium, details of the models for
the gas and electron densities, and potential contamination from point
sources.

The accuracy of this technique has continued to improve as
interferometric radio observations (e.g., Berkeley-Illinois-Maryland
Association, BIMA and Owens Valley Radio Observatory, OVRO) and ROSAT
and now Chandra X-ray data have become available. In a recent study by
Bonamente et al. (2006), new Chandra X-ray measurements for 38
clusters in the redshift range 0.14 $<$ z $<$ 0.89 have been
obtained. Combining these data with BIMA and OVRO data for these same
clusters, and performing a Markov Chain Monte Carlo analysis, these
authors find a value of H$_o$ = 76.9$^{+3.9 ~+10.0}_{-3.4 ~~-8.0}$ km
s$^{-1}$ Mpc$^{-1}$, under the assumption of hydrostatic
equilibrium. Relaxing the assumption of hydrostatic equilibrium, and
adopting an isothermal $\beta$ model, they find H$_o$ = 73.7$^{+4.6
~+9.5}_{-3.8 ~-7.6}$ km s$^{-1}$ Mpc$^{-1}$.

\subsection{Measurements of Anisotropies in the Cosmic Microwave
Background}

The prediction of acoustic oscillations in the cosmic microwave
background radiation (Peebles \& Yu 1970; Sunyaev \& Zel'dovich 1970)
and the subsequent measurement of these peaks (culminating most
recently in the five (Dunkley et al. 2009) and seven-year (Bennett et
al. 2010) measurements of WMAP, the Wilkinson Microwave Anisotropy
Probe) is one of the most successful chapters in the history of
cosmology.  A recent detailed review of the cosmic microwave
background is given in Hu \& Dodelson (2002).  The aim in this section
is simply to elucidate the importance of accurate measurements of the
Hubble constant in the context of measurements of the angular power
spectrum of CMB anisotropies, and the complementary nature of the
constraints provided.

The temperature correlations in the maps of the CMB can be described
by a set of spherical harmonics. A plot of the angular power spectrum
as a function of multipole moment, $l$, is shown in Figure 11.  This
spectrum can be naturally explained as a result of the tight coupling
between photons and baryons before recombination (where electrons and
protons combine to form neutral hydrogen), and a series of
oscillations are set up as gravity and radiation pressure act on the
cold dark matter and baryons. After recombination, photons free-stream
toward us. The position of the first peak in this diagram is a
projection of the sound horizon at the time of recombination, and
occurs at a scale of about 1 degree.

Although measurements of the CMB power spectrum can be
made to very high statistical precision, there are some nearly exact
degeneracies that limit the accuracy with which cosmological
parameters can be estimated  (e.g., Efstathiou \& Bond 1999). These
degeneracies impose severe limitations on estimates of curvature
and the Hubble constant derived from CMB anisotropy alone, and are
sometimes overlooked. Specifically, the value of H$_o$ is degenerate
with the value of $\Omega_\Lambda$ and $w$.  Different combinations of
the matter and energy densities and H$_o$ can produce identical CMB
anisotropy spectra.  Alternatively, an accurate independent
measurement of H$_o$ provides a means of constraining the values of
other cosmological parameters based on CMB anisotropy data.

The WMAP data provide strong evidence for the current standard
cosmological model with $\Omega_{matter}$ = 0.23, $\Omega_\Lambda$ =
0.73 (Spergel et al. 2003; Komatsu et al. 2010). A prior on H$_\circ$
can help to break some of the degeneracies in the CMB data.  The WMAP
data measure $\Omega_{matter} h^2$; assuming a flat universe, yields a
stronger constraint on the equation of state, -0.47 $,$ w $<$ 0.42
(95\% CL) (Komatsu et al. 2009) than WMAP data alone. Alternatively,
combining the WMAP-5 data with SNe~Ia and BAO data yields a value of
H$_0$ = 70.5 $\pm$ 1.3 km s$^{-1}$ Mpc$^{-1}$ (Komatsu et al. 2009),
still in excellent agreement with other methods.

\subsubsection{Measurements of Baryon Acoustic Oscillations in the
       Matter Power Spectrum}

Baryon acoustic oscillations (BAO) arise for the same underlying
physical reason as the peaks and valleys in the cosmic microwave
background spectrum: the sound waves that are excited in the hot
plasma owing to the competing effects of radiation pressure and
gravity at the surface of last scattering also leave an imprint on the
galaxy matter power spectrum. The two-point correlation function has a
peak on scales of 100 h$^{-1}$ Mpc (Eisenstein et al. 2005), which
provides a ``standard ruler'' for measuring the ratio of distances
between the surface of last scattering of the CMB (at z=1089) and a
given redshift.  Measurement of BAO in the matter power spectrum can
also help to break degeneracies in the CMB measurements. For example,
Percival et al. (2009) have combined the Sloan Digital Sky Survey
(SDSS) 7th data release with the Two-degree Field Galaxy Redshift
Survey (2dFGRS) to measure fluctuations in the matter power spectrum
at six redshift slices.  For $\Lambda CDM$ models, combining these
results with constraints for the baryon and cold dark matter
densities, $\Omega_bh^2$, and $\Omega_{CDM}h^2$ from WMAP 5, and data
for SNe~Ia, yields $\Omega_{matter}$ = 0.29 $\pm$ 0.02 and H$_o$ = 68
$\pm$ 2 km s$^{-1}$ Mpc$^{-1}$.

\section{AGE OF THE UNIVERSE}
\label{sec:age}

There are three independent ways of determining the age of the
Universe. The first is based on an assumed cosmological model and the
current expansion rate of the Universe. The second is based on models
of stellar evolution applied to the oldest stars in the Universe. The
third is based on measurements of the angular power spectrum of
temperature fluctuations in the CMB.  All three methods are completely
independent of each other, and so offer an important consistency
check. The kinematic age of the Universe is governed by the rate at
which the Universe is currently expanding, modified by the combined
extent to which gravity slows the expansion and dark energy causes it
to accelerate.

The time back to the Big Bang singularity depends upon $H_\circ$ and the
expansion history, which itself depends upon the composition of the
universe:
\begin{eqnarray}
t_o  = \int_o^\infty  {dz \over (1+z)H(z)}  = H_\circ^{-1}\int_o^\infty
{dz\over  (1+z)[\Omega_{matter}(1+z)^3 +\Omega_{DE}  (1+z)^{3(1+w)}  ]
^{1/2} }
\end{eqnarray}

\noindent
For a matter-dominated flat universe with no dark energy
($\Omega_{matter}$ = 1.0, $\Omega_{vacuum}$ = 0.0, the age is simply
2/3 of the Hubble time, or only 9.3 billion years for h = 0.7.

Not accounting for the presence of dark energy in the Universe leads
to an underestimate of its age. Before the discovery of dark energy,
an ``age controversy'' persisted for several decades: values of the
Hubble constant any larger than 40-50 km s$^{-1}$ Mpc$^{-1}$ appeared
to yield ages for the universe as a whole that were smaller than
stellar evolution calibrated ages of the oldest stars in the Milky
Way. For a universe with a Hubble constant of $73\,{\rm km\,sec^{-1}\,
Mpc^{-1}}$, with $\Omega_{matter}$ = 0.27 and $\Omega_{vacuum}$ =
0.73, the age is 13.3~Gyr.  Taking account of the systematic
uncertainties in $H_\circ$ alone, the uncertainty in the age of the
Universe is estimated to be about $\pm 0.8$~Gyr.

The most well-developed of the stellar chronometers employs the oldest
stars in globular clusters in the Milky Way (Krauss \& Chaboyer
2003). The largest uncertainty for this technique comes from
determination of the distances to the globular clusters. Recent,
detailed stellar evolution models when compared to observations of
globular clusters stars, yield a {\it lower limit} to their ages of
10.4 billion years (at the 95\% confidence level) with a best-fit age
of 12.6 Gyr.  Deriving the age for the Universe from the lower limit
requires allowing for additional time to form the globular clusters:
from theoretical considerations this is estimated to be about 0.8
billion years.  This age estimate for the Universe agrees well with
the expansion age.  Two other stellar chronometers:  the cooling of
the oldest white dwarf stars (for a recent review see Moehler \& Bono
2008) and nucleocosmochronology, the decay of radioactive isotopes
(Sneden et al. 2001),  yield similar ages.

The expansion age can also be determined from measurements of the CMB
anisotropy. $H_\circ$ cannot be measured directly from the CMB alone,
but the heights of the peaks in the CMB spectrum provide a constraint
on the product $\Omega_{matter} H_\circ^2$, and the position of the
peaks constrain the distance to the last-scattering surface. Assuming
a flat universe yields a consistent age, $t_o = 13.7\pm 0.13\,$Gyr
(Spergel et al. 2003; Komatsu et al. 2009), again in good agreement
with the other two techniques.

In summary, several methods of estimating the age of the universe are
now in good agreement, to within their quoted uncertainties, with a
value $t_o = 13.7 \pm 0.5\,$Gyr.

\section{WHY MEASURE  H$_o$ TO HIGHER ACCURACY?}
\label{sec:future}

The importance of pinning down $H_\circ$ has only grown with time: not
only does it set the scale for all cosmological distances and times,
but its accurate determination is also needed to take full advantage
of the increasingly precise measurements of other cosmological
quantities. The prospects for improving the accuracy of H$_o$ within
the next decade appear to be as exciting as those within the past
couple of decades. We highlight here near-term improvements to the
Cepheid-based extragalactic distance scale that will come from new
measurements of Cepheid parallaxes with GAIA and perhaps the Space
Interferometry Mission (SIM), {\it Spitzer} measurements of Cepheids
in the Milky Way, LMC, and other nearby galaxies, including NGC 4258,
{\it Spitzer} measurements of the Tully-Fisher relation and a new
calibration of the Type Ia supernova distance scale; and future
measurements of Cepheids with JWST. We describe how a more accurate
value of H$_o$, combined with other future measurements of large-scale
structure and CMB anisotropies (e.g., Planck), can be used to break
degeneracies and place stronger constraints on other cosmological
parameters including the equation of state for dark energy, the energy
density in cold dark matter, and the mass of neutrinos.

While measurements of CMB anisotropies have provided dramatic
confirmation of the standard concordance model, it is important to
keep in mind that the values for many quantities (e.g., $w$, H$_o$,
neutrino masses) are highly model-dependent, owing to the strong
degeneracies. A more accurate, independent measurement of H$_o$ is
critical for providing stronger limits on quantities such as $w$ and
neutrino masses.

\subsection{Constraints on Dark Energy}

As summarized by Hu (2005), a measurement of H$_o$ to the percent
level, in combination with CMB measurements with the statistical
precision of the Planck satellite offers one of the most precise
measurements of the equation of state at z $\sim$ 0.5. At first this
result appears counter-intuitive, since the CMB anisotropies result
from physical processes imprinted on the surface of last scattering at
z $\sim$ 1100. Alone they give very little information on dark energy,
which contributes most to the expansion at lower redshifts. However,
the sound horizon provides a reference standard ruler that can be used
to provide constraints on a number of parameters including dark energy
and neutrinos. The main deviations in the Hubble parameter, the
angular diameter distance, and the growth factor due to the dark
energy equation of state manifest themselves as variations in the
local Hubble constant. In Figure 12, we show the strong degeneracy
between the equation of state and the value of H$_o$.  This figure is
based on a forecast of the precision that will be available with
measurements of CMB fluctuations from the Planck satellite.  Improved
accuracy in the measurement of H$_o$ will be critical for constraining
the equation of state for dark energy from CMB data.

\subsection{Constraints on the Neutrino Mass}

Improved accuracy in the measurement of H$_o$ will have a significant
effect in placing constraints on the neutrino mass from measurements
of CMB anisotropies. Detailed reviews of the subject can be found in
Dolgov (1996), Crotty, Lesgourges \& Pastor (2004) and Hannestad
(2006).  Briefly, massive neutrinos contribute to the overall matter
density of the Universe through which they have an impact on the
growth of structure; the larger the neutrino mass, the more
free-streaming effects dampen the growth of structure on small
scales. The density in neutrinos is related to the number of massive
neutrinos, $N_{eff}$, and the neutrino mass, $m_\nu$, by: $\Delta_\nu
h^2 = N_{eff} m_\nu $ / 94 eV.  From neutrino oscillation experiments,
a measurement of the difference in mass squared, $\Delta m^2 \sim$
0.002 (eV)$^2$ is obtained.

In the context of the standard cosmological model, cosmological
observations can constrain the number of neutrino species and the
absolute mass scale. Massive neutrinos have a measurable effect on the
cosmic microwave background spectrum: the relative height of the
acoustic peaks decrease with increasing $m_\nu$ and the positions of
the peaks shift to higher multipole values. The WMAP 5-year data
provided evidence, for the first time, for a non-zero neutrino
background from CMB data alone, with $\sum m_\nu < $ 1.3 eV (95\% CL)
(Dunkley et al. 2009). Combining the CMB data with results from SNe~Ia
and baryon acoustic oscillations, results in a bound of $\sum m_\nu <
$ 0.58 eV (95\% CL) (Komatsu et al. 2010), reaching close to the range
implied by the neutrino oscillation experiments. Future forecasts with
Planck data suggest that an order of magnitude increase in accuracy
may be feasible.  One of the biggest limitations to determining the
neutrino mass from the CMB power spectrum results from a strong
degeneracy between the neutrino mass and the Hubble constant (Komatsu
et al. 2009).  As $H_ \circ$ increases, the neutrino mass becomes
smaller (see Figure 13).  An accuracy in H$_o$ to 2-3\% percent,
combined with Planck data (for the standard cosmological model) will
provide an order of magnitude improved precision on the neutrino mass.

\subsection{Measuring~~H$_o$~~to~~$\pm$2\%}
\label{sec:futureceph}

Accuracy in measurement of H$_\circ$ has improved signficantly with
the measurement of HST Galactic Cepheid parallaxes and HST measurement
of Cepheid distances to SNe Ia hosts, as described in \S
\ref{sec:plcalibration} and \S \ref{sec:snh0}, respectively.  Future
improvements will come with further HST WFC3 and {\it Spitzer}
observations of Cepheids.  At 3.6 and 4.5$\mu$m the effects of
extinction are a factor of $\sim$20 smaller in comparison to optical
wavelengths. In addition, in the mid-infrared, the surface brightness
of Cepheids is insensitive to temperature. The amplitudes of the
Cepheids are therefore smaller and due to radius variations alone.
The Leavitt Law in the mid-IR then becomes almost equal to the
Period-Radius relation. From archival {\it Spitzer} data, the
mid-infrared Leavitt Law has been shown to have very small dispersion
(Freedman et al. 2008; Madore et al. 2009).  Furthermore, metallicity
effects are expected to be small in the mid infrared, and {\it
Spitzer} offers an opportunity to test this expectation
empirically. The calibration can be carried out using Spitzer alone,
once again eliminating cross-calibration uncertainties. A new program
aimed at addressing remaining systematic errors in the Cepheid
distance scale is the Carnegie Hubble Program (CHP: Freedman 2009).

The CHP will measure the distances to 39 Galactic Cepheids (15 of them
in anticipation of the GAIA satellite), 92 well-observed Cepheids in
the LMC, several Local Group galaxies containing known Cepheids (M31,
M33, IC~1613, NGC~6822, NGC~3109, Sextans~A, Sextans~B and WLM), more
distant galaxies with known Cepheids including NGC~2403 (2.5~Mpc),
Sculptor Group galaxies NGC~ 300, NGC 247 (3.7~Mpc), Cen~A (3.5~Mpc)
and M83 (4.5~Mpc), as well as the maser galaxy NGC 4258 (at 7.2~Mpc).
It will measure the distances to 545 galaxies in 35 clusters with
measured Tully-Fisher distances, which can then be calibrated with
Cepheids as shown in Figure 7.  Over 50 galaxies with SNe~Ia distances
measured by Folatelli et al. (2009) will also be observed as part of
this program, allowing a determination of H$_o$ with this calibration
well into the far-field Hubble flow.

As discussed earlier, the expected uncertainties from the CHP are
shown in Table 2.  Re-observing the known Cepheids in more distant
galaxies will require the aperture, sensitivity and resolution of
JWST.  With {\it Spitzer}, it will be possible to decrease the
uncertainties in the Cepheid distance scale to the (3-4\%) level, with
an application of a new mid-IR Tully-Fisher relation and a {\it
Spitzer} Cepheid calibration of Type~Ia SNe.  It is expected that
future JWST measurements will bring the uncertainties to $\pm$2\% with
a more firm calibration of SNe~Ia.

\section{FUTURE IMPROVEMENTS}

We summarize here the steps toward measuring the Hubble constant to a
few percent accuracy. Most of these measurements should be feasible
within the next decade.

1. Mid-infrared Galactic Cepheid parallax calibration with {\it
  Spitzer} and GAIA.

2. Mid-infrared calibrations of Galactic and nearby Cepheid galaxies
and the infrared Tully-Fisher relation with {\it Spitzer} and JWST.

3. Increased numbers of maser distances.

4. Larger samples and improved systematics and modeling of strong
gravitational lenses and Sunyaev-Zel-dovich clusters.

5. Higher-frequency, greater sensitivity, higher angular resolution
measurements of the CMB angular power spectrum with Planck.

6.  Measurements of baryon acoustic oscillations at a range of
redshifts (e.g.,

BOSS   [http://cosmology.lbl.gov/BOSS/],
HETDEX  [http://hetdex.org/],
WiggleZ [http://wigglez.swin.edu.au/site/]),
JDEM [http://jdem.gsfc.nasa.gov/],
SKA  [http://www.skatelescope.org/],
DES  [http://www.darkenergysurvey.org/],
PanStarrs [http://pan-starrs.ifa.hawaii.edu/public/],
LSST [http://www.lsst.org/lsst]).

7.  Beyond 2020, detection of gravitational radiation from inspiraling
massive black holes with LISA. Coupled with identification with an
electromagnetic source and therefore a redshift, this method offers,
in principle, a 1\% measurement of H$_0$.

\vfill\eject
\section{SUMMARY POINTS}

(1) Several nearby distance determination methods are now available
that are of high precision, having independent systematics. These
include Cepheid variables, the tip of the red giant branch (TRGB)
stars, and the geometrically determined distances to maser galaxies.

(2) The Cepheid Period-Luminosity relation (Leavitt Law) now has an
absolute calibration based on HST trigonometric parallaxes for
Galactic Cepheids. This calibration and its application at
near-infrared wavelengths significantly reduces two of the four
leading systematic errors previously limiting the accuracy of the
Cepheid-based distance scale: zero-point calibration and metallicity
effects.

(3) The maser galaxy distances, TRGB distances and Cepheid distances
agree to high precision at the one common point of contact where they
can each be simultaneously intercompared, the maser galaxy NGC 4258,
at a distance of 7.2  Mpc.

(4) Galactic Cepheid parallax and NGC 4258 maser calibrations of the
distance to the LMC agree very well.  Based on these measurements and
other independent measurements, we adopt a true, metallicity-corrected
distance modulus to the LMC of 18.39 $\pm$ 0.06 mag.

(5) HST optical and near-infrared observations of Cepheids in SNe Ia
galaxies calibrated by the maser galaxy, NGC 4258, have decreased
systematics due to calibration, metallicity and reddening in  the SNe
Ia distance scale, and increased the number of well-observed SN
calibrators to six.

(6) The current calibration of the Cepheid and maser extragalactic
distance scales agree to within their quoted errors, yielding a value
of $H_\circ$ = 73 $\pm$ 2 (random) $\pm$ 4 (systematic)
km~s$^{-1}$~Mpc$^{-1}$.

(7) Within a concordance cosmology (that is, $\Omega_{matter}$ = 0.27 and
$\Omega_{vacuum}$ = 0.73) the current value of the Hubble constant
gives an age for the Universe of 13.3 $\pm$ 0.8~Gyr. Several
independent methods (globular cluster ages, white dwarf cooling ages,
CMB anisotropies within a concordance model) all yield values in good
agreement with the expansion age.

(8) Further reductions of the known systematics in the extragalactic
distance scale are anticipated using HST, {\it Spitzer}, GAIA and
JWST. A factor of two decrease in the currently identified systematic
errors is within reach, and an uncertainty of 2\% in the Hubble
constant is a realistic goal for the next decade.

(9) A Hubble constant measurement to a few percent accuracy, in
combination with measurements of anisotropies in the cosmic microwave
background from Planck, will yield valuable constraints on many other
cosmological parameters, including the equation of state for dark
energy, the mass of neutrinos, and the curvature of the universe.

\section{DISCLOSURE STATEMENT}
The authors are not aware of any potential biases that might be
perceived as affecting the objectivity of this review.

\section{ACKNOWLEDGEMENTS}

We thank the Carnegie Institution for Science, which, through its
support of astronomical facilities and research for over a century,
enabled the original discovery of the expansion of the Universe, as
well as continued efforts to measure accurately the expansion of the
universe over cosmic time.  Many dedicated individuals have made
contributions to solving the problems encountered in accurately
calibrating the extragalactic distance scale; it has been a community
effort spanning the better part of a century, but it remained a
problem that could not have been solved without NASA and its vision in
supporting space astrophysics.  We thank Chris Burns for producing
Figure 12. We also thank Fritz Benedict, Laura Ferrarese, Robert
Kirshner, James Moran, Jeremy Mould, and Adam Riess for their many
constructive suggestions on an earlier draft of this article. 
And finally, our sincere thanks to John Kormendy, our  ARAA Editor, for his
patience and his insightful and helpful comments on the final version of
this review as it went to press.

\vfill\eject

\pagebreak

\begin{table}
\def~{\hphantom{0}}
\caption{Galactic Cepheids with Geometric Parallaxes}\label{tab1}
\begin{tabular}{@{}llcccl@{}}
\toprule
Cepheid & P(days) & logP & $\mu$~(mag.) & $\sigma$~(\%) & Distance
(pc)\\
\colrule
RT Aur      & ~3.728  & 0.572  &  8.15 & ~7.9 & 427 \\
T Vul       & ~4.435  & 0.647  &  8.73 & 12.1 & 557 \\
FF Aql      & ~4.471  & 0.650  &  7.79 & ~6.4 & 361 \\
$\delta$~Cep  & ~5.366  & 0.730  &  7.19 & ~4.0 & 274 \\
Y Sgr       & ~5.773  & 0.761  &  8.51 & 13.6 & 504 \\
X Sgr       & ~7.013  & 0.846  &  7.64 & ~6.0 & 337 \\
W Sgr       & ~7.595  & 0.881  &  8.31 & ~8.8 & 459 \\
$\beta$~Dor   & ~9.842  & 0.993  &  7.50 & ~5.1 & 316 \\
$\zeta$~Gem   & 10.151  & 1.007  &  7.81 & ~6.5 & 365 \\
l Car       & 35.551  & 1.551  &  8.56 & ~9.9 & 515 \\
\botrule
\end{tabular}
\end{table}

\pagebreak

\begin{table}
\caption{Systematics Error Budget on $H_\circ$: Past, Present, Future}
\label{tab3}
\begin{tabular}{lllll}
\hline \\
Known& Key Project     & Revisions  & Anticipated  & Basis  \\
Systematics &  (2001)  & (2007/2009)   & Spitzer/JWST  &   \\
\hline \\
(1) Cepheid Zero Point & $\pm$0.12~mag &$\pm$0.06~mag & $\pm$0.03~mag
& Galactic Parallaxes \\
(2) Metallicity   &    $\pm$0.10~mag   &$\pm$0.05~mag & $\pm$0.02~mag
& IR + Models \\
(3) Reddening    &     $\pm$0.05~mag   &$\pm$0.03~mag & $\pm$0.01~mag
& IR 20-30x Reduced \\
(4) Transformations &  $\pm$0.05~mag   &$\pm$0.03~mag & $\pm$0.02~mag
& Flight Magnitudes \\
& & & \\
{\bf Final Uncertainty}&  $\pm$0.20~mag &$\pm$0.09~mag  & $\pm
$0.04~mag&  Added in Quadrature \\
Percentage Error on $H_\circ$ & $\pm$10\% & $\pm$5\%    &  $\pm$2\% &
Distances\\
& & & \\
\hline \\
\end{tabular}
{\footnotesize ~~Revisions (Column 2) incorporating the recent work of
Benedict et al. (2007) and Riess et al. (2009a).}
\end{table}

\pagebreak

\begin{figure}
\centerline{\psfig{figure=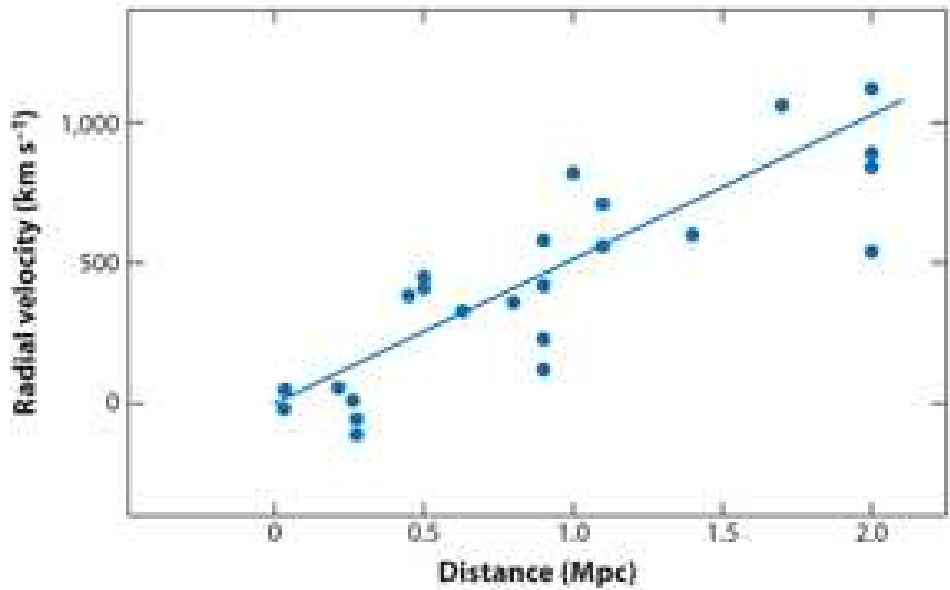,height=20pc,angle=00}}
\caption{From Hubble (1929a): radial velocities, corrected for solar
motion, plotted versus distances estimated from stars and mean
luminosities of galaxies in clusters. The solid dots and line represent
the solution for solar motion using individual galaxies.  Hubble
wrote: {\it ``The outstanding feature, however, is the possibility
that the velocity-distance relation may represent the de Sitter
effect, and hence that numerical data may be introduced into
discussions of the general curvature of space.}''}
\label{fig1}
\end{figure}

\clearpage
\begin{figure}
\centerline{\psfig{figure=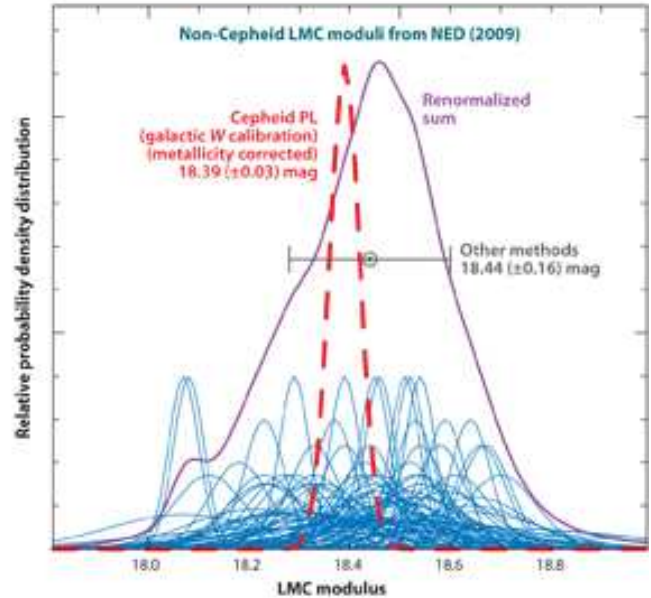,height=20pc,angle=00}}
\caption{The cumulative probability density distribution of 180
distance modulus estimates to the LMC culled from the recent
literature, provided by NED. Individual estimates are shown by
unit-area gaussians with a dispersion set to their quoted statistical
errors. The thin solid line represents the renormalized sum of those
gaussians.  The thick broken line represents the value of 18.39~mag
and a systematic error of $\pm$0.03 mag for the true (Wesenheit)
distance modulus to the LMC, based on the Galactic parallax
calibration for Cepheids and corrected for metallicity by -0.05~mag.
For comparison the median value of the published, non-Cepheid distance
moduli is 18.44$\pm$0.16~mag (shown as the circled point and error
bar); the mode of the non-Cepheid moduli is 18.47~mag. The Cepheid
value is statistically indistinguishable from this highly
heterogeneous, but fairly complete, set of independently published
determinations.}
\label{fig2}
\end{figure}

\begin{figure}
\centerline{\psfig{figure=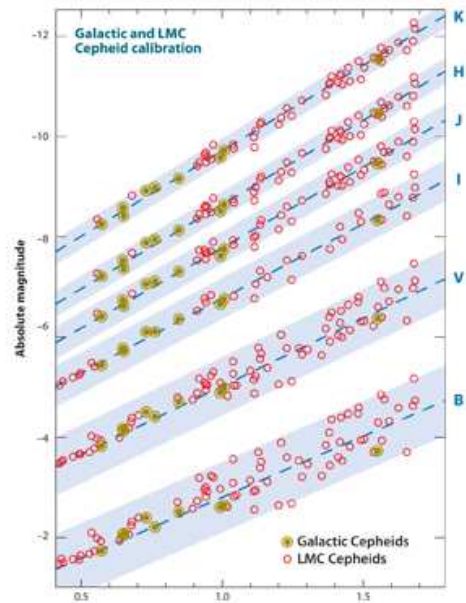,height=20pc,angle=00}}
\caption{Composite multiwavelength Period-Luminosity relations
(Leavitt Laws) for Galactic (circled filled dots) and LMC (open
circles) Cepheids from the optical (BVI) through the near-infrared
(JHK).  There is a  monotonic increase in the slope, coupled with a
dramatic decrease in total dispersion of the PL relations as one goes
to longer and longer wavelengths.}
\label{fig3}
\end{figure}

\begin{figure}
\centerline{\psfig{figure=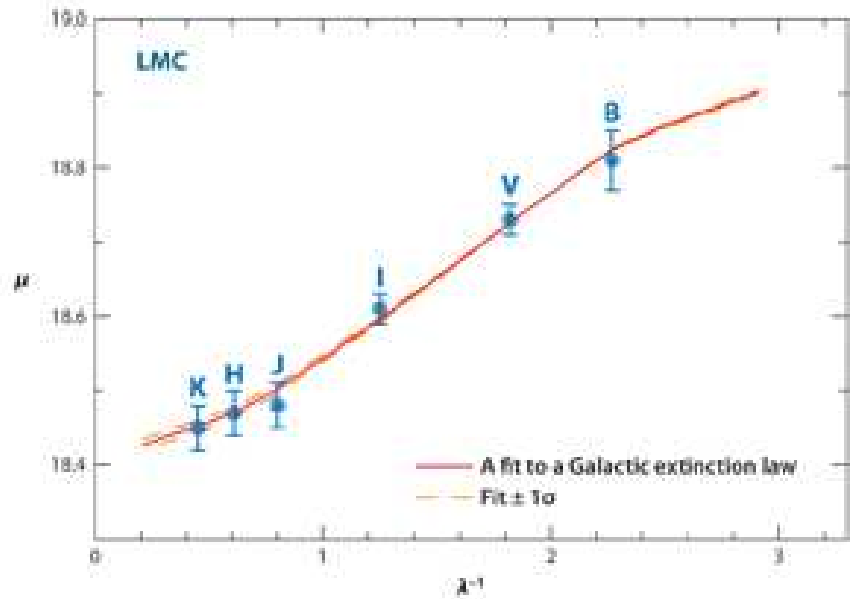,height=20pc,angle=00}}
\caption{Standard extinction-curve fit to six multiwavength (BVIJHK)
apparent distance moduli to the LMC scaled to the HST Galactic
parallax sample (Benedict et al. 2007). The minimized-$\chi^2$ scaled
fit gives a true distance modulus (intercept) of 18.40$\pm$0.01~mag,
uncorrected for metallicity, and a total line-of-sight color excess
(slope) of E(B-V) = 0.10 mag. }
\label{fig4}
\end{figure}

\begin{figure}
\centerline{\psfig{figure=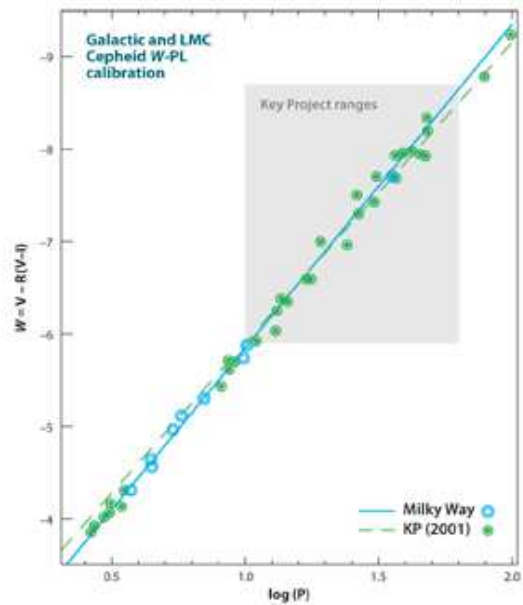,height=20pc,angle=00}}
\caption{The reddening-free VI Wesenheit PL relation showing the
combined data for Galactic Cepheids having individually-determined
trigonometric parallaxes (circled dots) and Large Magellanic Cloud
Cepheids (open circles) brought into coincidence with the Galactic
calibration after an offset of 18.44~mag between their apparent
magnitudes. The solid line is a fit to the combined data. The dashed
line is the calibration used by Freedman et al. (2001) at the
conclusion of the Key Project. The inner bounding box shows the period
and luminosity range used by the Key Project to determine
extragalactic distances. The correspondence between the two
calibrations is very close, but it should be noted that the Galactic
calibration is for Galactic metallicity. }
\label{fig5}
\end{figure}

\begin{figure}
\centerline{\psfig{figure=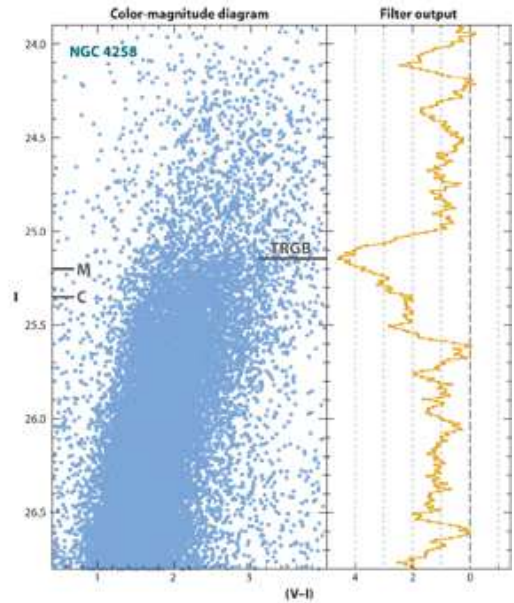,height=20pc,angle=00}}
\caption{An example of the detection and measurement of the
discontinuity in the observed luminosity function for red giant branch
stars in the halo of the maser galaxy NGC~4258 (Mager, Madore \&
Freedman 2008). The color-magnitude diagram on the left has been
adjusted for metallicity such that the TRGB is found at the same
apparent magnitude independent of color/metallicity of the stars at
the tip.  The right panel shows the output of an edge-detection
(modified Sobel) filter whose peak response indicates the TRGB
magnitude and whose width is used as a measure of the random error on
the detection.  }
\label{fig6}
\end{figure}

\begin{figure*}
\centerline{\psfig{figure=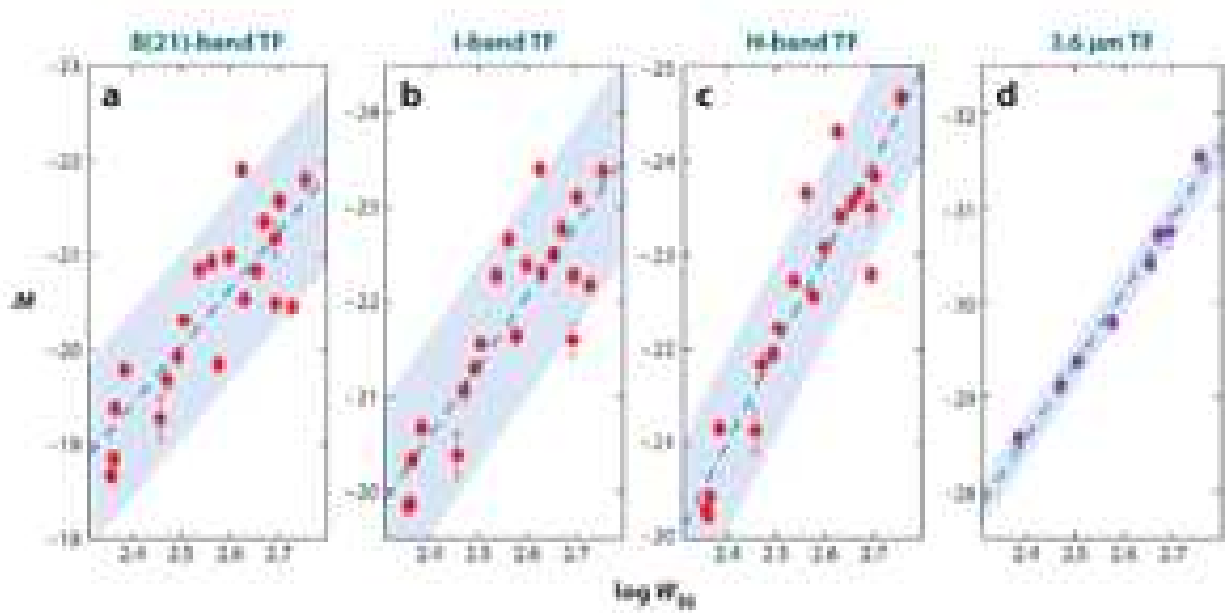,height=20pc,angle=00}}
\caption{\small Multi-wavelength Tully-Fisher relations. The three
left panels show the B,I and H-band TF relations for all of the
galaxies calibrated with independently-measured Cepheid moduli from
the HST Key Project.  The right-hand panel shows the TF relation for
the subset of galaxies drawn from the Key Project calibrators that
also have published 3.6$\mu$m total magnitudes. The significantly
reduced dispersion in the mid-infrared dataset is impressive; however,
a larger sample of calibrators is needed to confirm the scatter and
slope of the relation at this wavelength.}
\label{fig7}
\end{figure*}

\begin{figure}
\centerline{\psfig{figure=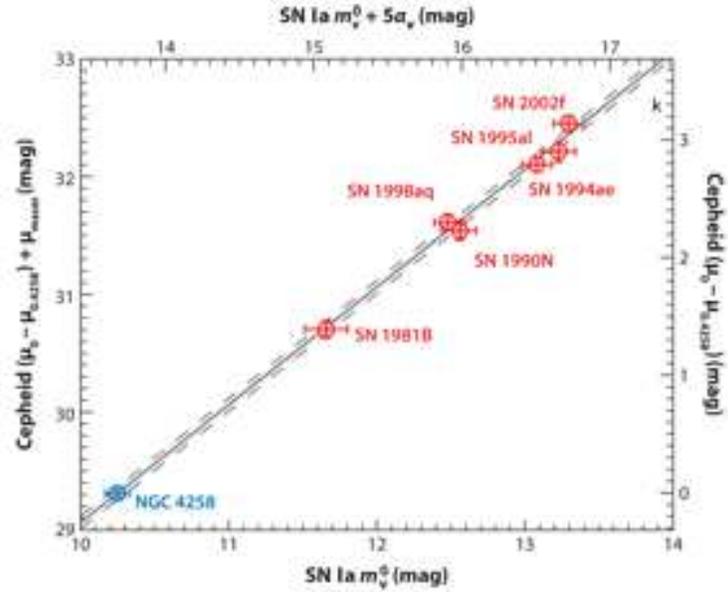,height=20pc,angle=00}}
\caption{A comparison of Cepheid and SNe Ia distances (red points), as
described in Riess et al. (2009a). The calibrating galaxy, NGC 4258,
is added in blue.}
\label{fig8}
\end{figure}

\begin{figure}
\centerline{\psfig{figure=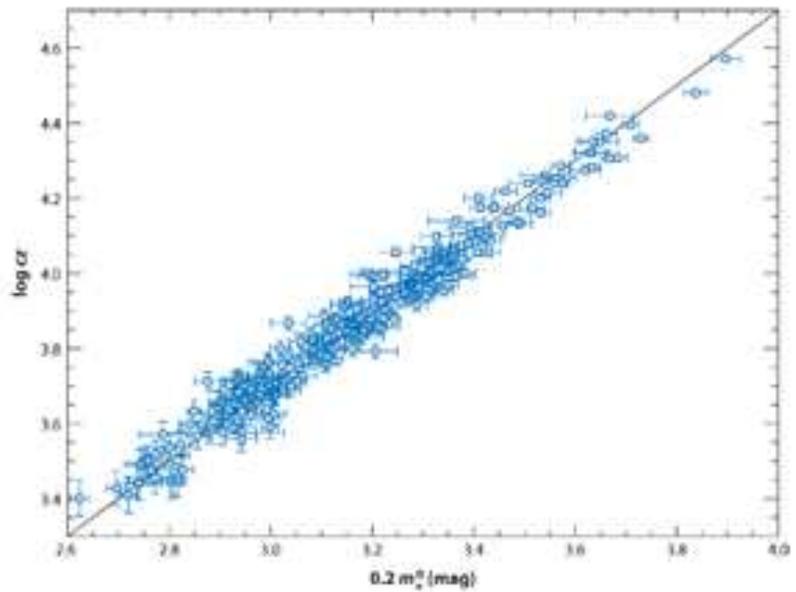,height=20pc,angle=00}}
\caption{Supernova Hubble diagram based on 240 supernovae with z $<$
    0.1. The sample is from Hicken et al. (2009), and have been used by
    Riess et al. (2009a) for their determination of H$_o$.}
\label{fig9}
\end{figure}

\begin{figure}
\centerline{\psfig{figure=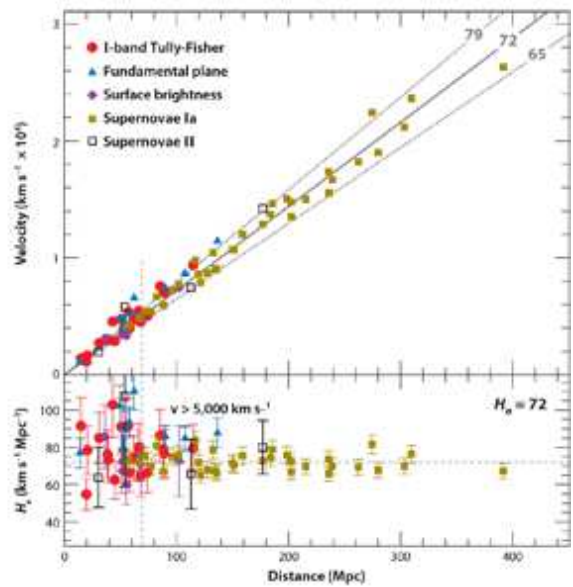,height=20pc,angle=00}}
\caption{Graphical results of the HST Key Project (Freedman et
al. 2001).  Top Panel: The Hubble diagram of distance vs. velocity for
secondary distance indicators calibrated by Cepheids. Velocities are
corrected using the nearby flow model of Mould et al. (2000). Squares:
Type Ia supernovae; filled circles: Tully-Fisher clusters (I-band
observations); triangles: fundamental plane clusters; diamonds:
surface brightness fluctuation galaxies; open squares: Type II
supernovae. A slope of $H_\circ$ = 72 $\pm$ 7~ km~s$^{-1}$~Mpc$^{-1}$ is
shown. Beyond 5000 km/s (vertical line), both numerical simulations
and observations suggest that the effects of peculiar motions are
small. The Type Ia supernovae extend to about 30,000~km/s, and the
Tully-Fisher and fundamental plane clusters extend to velocities of
about 9,000 and 15,000~km/s, respectively. However, the current limit
for surface brightness fluctuations is about 5,000~km/s. Bottom Panel:
The galaxy-by-galaxy values of $H_\circ$ as a function of distance.}
\label{fig10}
\end{figure}

\begin{figure}
\centerline{\psfig{figure=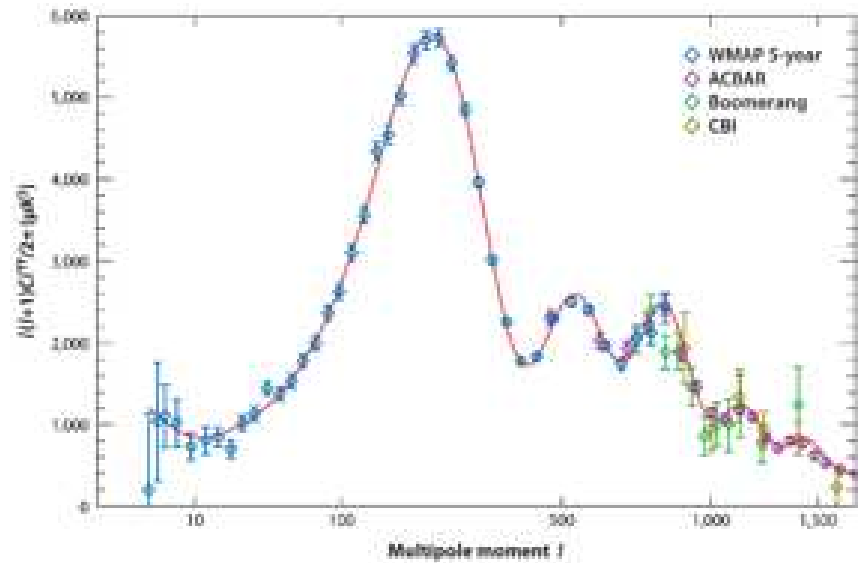,height=20pc,angle=00}}
\caption{The {\it WMAP} 5-year ``temperature angular power spectrum''
(Nolta et al. 2009) incorporating other recent results from the {\it
ACBAR} (Reichardt et al. 2008, purple), {\it Boomerang} (Jones et
al. 2006, green), and {\it CBI} (Readhead et al. 2004, red)
experiments. The red curve is the best-fit {\it CDM} model to the {\it
WMAP} data.}
\label{fig11}
\end{figure}

\begin{figure}
\centerline{\psfig{figure=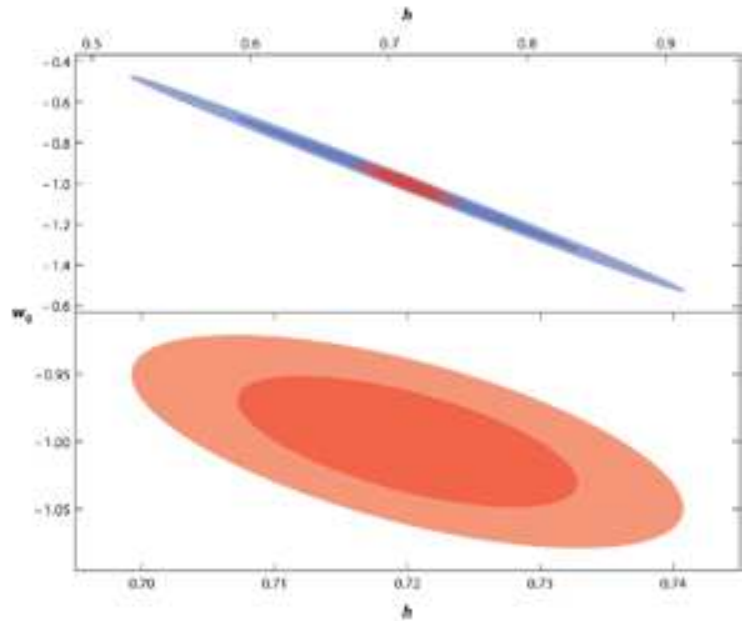,height=20pc,angle=00}}
\caption{Upper Panel: A plot illustrating the degeneracies of w$_o$
    with h = (H$_o$ / 100) assuming the statistical uncertainties
    expected for the Planck satellite, assuming a flat universe
    ($\Omega_k$ = 0), and constant dark energy (w$_a$ = 0). The plot
    uses the Planck Fisher matrix from the DETFast software package
    (Albrecht et al. 2006) The outer blue contours show the 68\% and
    95\% confidence intervals from the H$_o$ Key Project (h = 0.72
    $\pm$ 0.08), and the inner red contours show the case for a 2\%
    uncertainty in H$_o$.  Improved precision in H$_o$ will allow an
    accurate measurement of w from the CMB, independently of other
    methods. Lower Panel: same as above, adding in constraints from
    Stage III supernovae and baryon acoustic oscillation experiments,
    as described in Albrecht et al.}
\label{fig12}
\end{figure}

\begin{figure}
\centerline{\psfig{figure=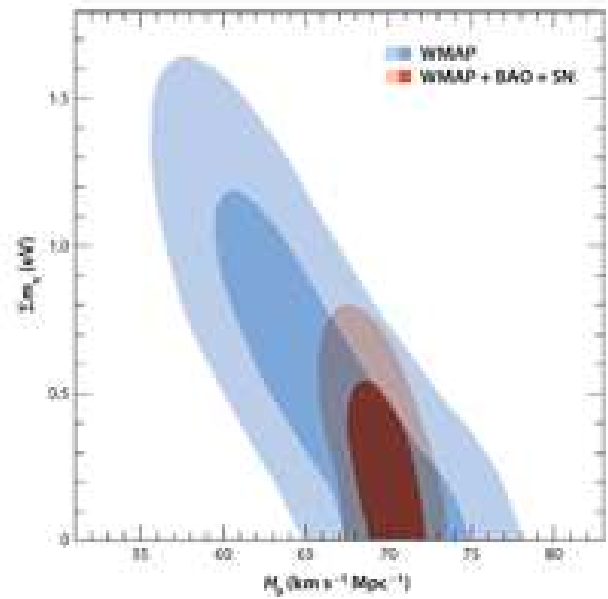,height=20pc,angle=00}}
\caption{WMAP5 data showing the degeneracy between the sum of neutrino
    masses and H$_o$ (Figure 17, Komatsu et al. 2009). The blue contours
    show the WMAP5 data only (68\% and 95\% CL; the red contours include
    BAO and SNe~Ia data.}
\label{fig13}
\end{figure}

\end{document}